\documentclass[aps,prl,reprint,superscriptaddress,amsmath,amssymb,a4paper,twocolumn,longbibliography,footinbib]{revtex4-1}

\usepackage[utf8]{inputenc}
\usepackage{amsmath}
\usepackage{amssymb,amsfonts,latexsym}
\usepackage{bm}
\usepackage[mathcal]{euscript}
\usepackage{graphicx}
\usepackage{epsfig}
\usepackage{color}
\usepackage{pifont,wasysym,marvosym}
\usepackage{textcomp}
\usepackage{comment}
\usepackage{epstopdf}
\usepackage{dcolumn}
\usepackage{hyperref}
\definecolor{bleuf}{rgb}{0,0.44,0.72}

\newcommand{\rev}[1]{\textcolor{black}{#1}}

\begin{document}
\title{Self-oscillation and Synchronisation Transitions in Elasto-Active Structures}

\author{Ellen Zheng}
\affiliation{Institute of Physics, Universiteit van Amsterdam, Science Park 904, 1098 XH Amsterdam, The Netherlands}
\author{Martin Brandenbourger}
\affiliation{Institute of Physics, Universiteit van Amsterdam, Science Park 904, 1098 XH Amsterdam, The Netherlands}
\author{Louis Robinet}
\author{Peter Schall}
\affiliation{Institute of Physics, Universiteit van Amsterdam, Science Park 904, 1098 XH Amsterdam, The Netherlands}
\author{Edan Lerner}
\affiliation{Institute of Physics, Universiteit van Amsterdam, Science Park 904, 1098 XH Amsterdam, The Netherlands}
\author{Corentin Coulais}
\affiliation{Institute of Physics, Universiteit van Amsterdam, Science Park 904, 1098 XH Amsterdam, The Netherlands}

\begin{abstract}
The interplay between activity and elasticity often found in active and living systems triggers a plethora of autonomous behaviors ranging from self-assembly and collective motion to actuation. Amongst these, spontaneous self-oscillations of mechanical structures is perhaps the simplest and most wide-spread type of non-equilibrium phenomenon. Yet, we lack experimental model systems to investigate the various dynamical phenomena that may appear.
\rev{Here, we introduce a centimeter-sized model system for one-dimensional elasto-active structures.}
\rev{We show that such structures exhibit flagellar motion when pinned at one end, self-snapping when pinned at two ends, and synchronization when coupled together with a sufficiently stiff link.}
%
\rev{We further demonstrate that these transitions can be described quantitatively by simple models of coupled pendula with follower forces.} 
%
%
Beyond the canonical case considered here, we anticipate our work to open avenues for the understanding and design of the self-organisation and response of active biological and synthetic solids, e.g. in higher dimensions and for more intricate geometries.
\end{abstract}

\maketitle

\textit{Introduction.} --- 
Active matter systems exhibit exceptional collective, non-equilibrium properties resulting in anomalous dynamical and self-organizing behaviours that challenge conventional laws of statistical mechanics~\cite{PhysRevE.89.062316,RevModPhys.85.1143, ramaswamy2017active,RevModPhys.88.045006,DeseignePRL2010,PhysRevLett.110.208001, bricard2013emergence,dauchotPRLsinglebug,VICSEK201271,baskaran2009statistical}. 
While researches have extensively focused on active fluids~\cite{saintillan2013active,RevModPhys.85.1143}---which consist of collections of individual active particles with no particular geometry~\cite{REIMANN200257, paxton2004catalytic,howse2007self,lauga2009hydrodynamics, poon2013clarkia},
active solids---which have a well defined reference state and hence exhibit elastic rather than viscous properties at long timescales~\cite{hawkins2014stress,Brandenbourger_NatComm2019,Scheibner_NatPhys2020}---have been much less studied, despite their potential in mimicking living matter and forming novel active materials~\cite{Brandenbourger_NatComm2019,Scheibner_NatPhys2020,Goldman_blobspnas2021}.

\begin{figure}[b!]
 \begin{center}
  \includegraphics[width=\columnwidth,clip,trim=0cm 0cm 0cm 0cm]{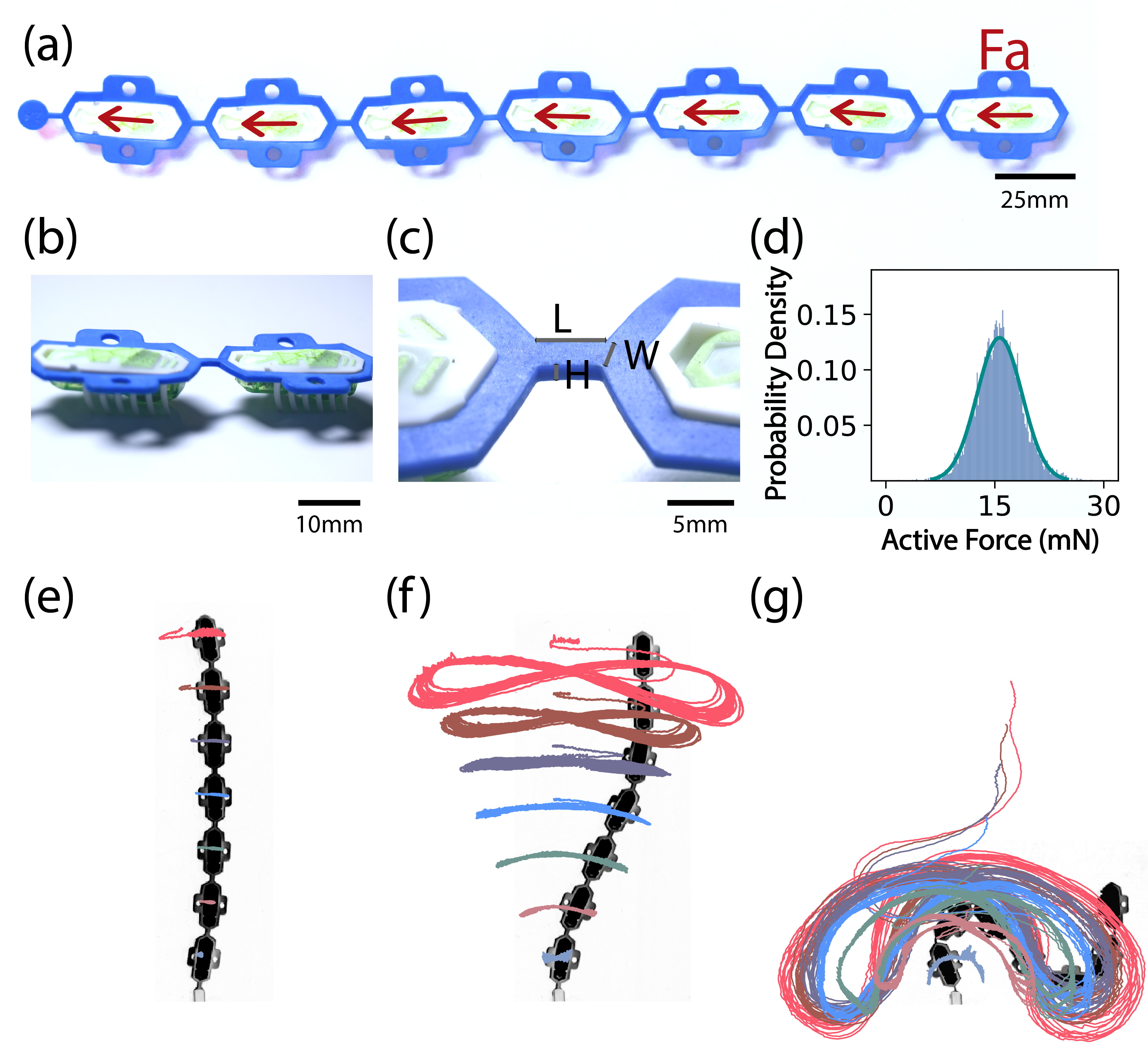}
 \end{center}
 \caption{Emergence of Self Oscillations in elasto-active chains.
 (a) Configurations of 7 active particles connected by a flexible rubber chain (b) zoomed in details of two active particles unveiling the design of the microbot (c) a close-up of the linkage between each particle with width (W), thickness (H) and length (L). (d) Histogram of the active force measurements conducted at 0.05 mm/min. (e)-(f) Snapshots of the trajectories of the active particles showing the oscillations changed from self-amplified to overdamped with W = 5 mm, 4.4mm and 2mm corresponding to elasto-active number $\sigma$ = 0.17, 0.21 and 0.80 respectively. \rev{See also Supplementary Videos.}}
 \label{fig:ae}
\end{figure}


Among all kinds of mechanical properties of active solids, self-oscillations are vital for biological systems such as flagella and cilia~\cite{machin1958wave,brokaw1975molecular,camalet1999self} and offer the prospect of autonomous mechanical behaviors in designer materials~\cite{woodhouse_PRL,Brandenbourger_NatComm2019,Scheibner_NatPhys2020}. It is now well established that one-dimensional active chains exhibit flagellar motion: on the one hand, experimental studies have reported self-oscillatory behavior and synchronization in biological and colloidal systems~\cite{nishiguchi2018flagellar,wan2014lag,PhysRevX.11.031051, brumley2014flagellar,goldstein2011emergence,manna2017colloidal}; on the other hand,  theoretical and numerical studies have suggested that self-oscillations 
emerge from the competition between activity and elasticity~\cite{hilfinger2008chirality,brokaw1975molecular,camalet1999self,hilfinger2009nonlinear,machin1958wave,fu2008beating,gladrow2017nonequilibrium,chelakkot2014flagellar,quaranta2015hydrodynamics,wan2016coordinated,niedermayer2008synchronization,guirao2007spontaneous,jayaraman2012autonomous, bennett2013emergent, vilfan2006hydrodynamic, wan2014lag,DeCanio_JRSI2017}. Despite these advances, there are as of yet few model experimental platforms in which the predicted bifurcation scenarios that lead to self-oscillations and synchronization can be verified. 


Here, to investigate dynamical transitions in elasto-active solids, we construct the experimental setup for a simplest form of active solids by elastically constraining centimeter-sized active particles in one-dimensional chains that can freely oscillate in the 2d plane. By controlling the elasticity of such structures, we uncover the nature of the transition to self-oscillations and synchronisation. 
We find the transition to \rev{flagellar and self-snapping motion} is governed by 
a nonlinear feedback between the direction of the active forces and the nonlinear elastic deflections. 
We find that synchronization between two elasto-active chains is mediated by elastically driven alignment, in contrast with active fluids. \rev{Although our proposed experimental platform is macroscopic, it might nonetheless help to advance our understanding of elasto-active instabilities that occur at the smaller scale in biological solids. We further envision that it will provide design guidelines for autonomous behaviors in active  solids~\cite{Baconnier_Arxiv2021,Brandenbourger_arXiv2021}.} 

\begin{figure}[t!]
 \begin{center}
  \includegraphics[width=.99\linewidth,clip,trim=0cm 0cm 0cm 0cm]{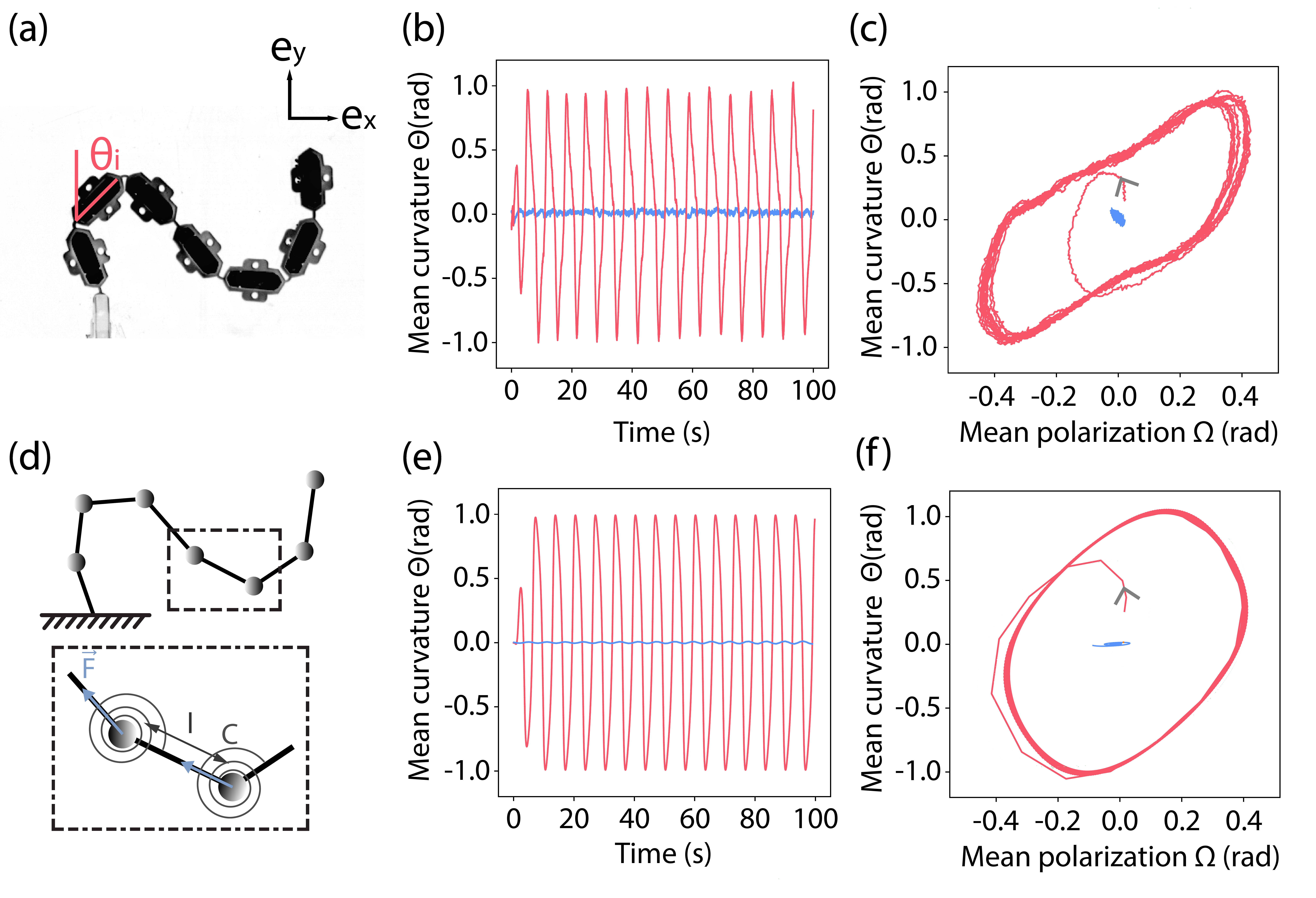}
 \end{center}
 \caption{
 Characterisation of the dynamics of elasto-active chains.
 (a) Snapshot of the elasto-active chain with W=17.7mm during its self-oscillation. (b) Time series of the angle between the first and last particle (mean polarisation) was one of the parameters we chose to characterize the system, blue and orange represent the $\sigma = 0.695$ and $\sigma=0.166$ chain respectively. With the average of tangential angles of the particles (mean curvature) being the other parameter, plotting mean polarisation ($\Theta$) against mean curvature ($\Omega$) gives (c) limit cycle showing that active forces balance with dissipation towards stable self-oscillation. (d) Schematics of an elasto-active chain of 7 pendulums with active forces. (e)-(f) Simulation results at $\sigma$=0.8 showing good agreement with the experimental results. }
\label{fig:be}
\end{figure}

\textit{Experimental design of active chains.} --- Our system consists of $N=7$ $5$cm commercial self-propelled microbots (Hexbug Nano v2) \cite{Christensen2016,dauchotPRLsinglebug} elastically coupled by a laser-cut silicon rubber chain pinned at one end as shown in FIG.\ref{fig:ae}(a,b)\footnote{\rev{The precise number of particles $N$, which does not affect the bifurcation phenomena, was chosen because of experimental considerations such as manufacturing and visualization. In addition, we also ran an additional experiment $N=17$, which displays the same self-oscillating pattern (see SI). Finally, we performed a linear stability analysis in the SI, which also demonstrates that the unstable mode does not depend on system size.}}.
By tuning the width of the connection ($W$ in FIG.\ref{fig:ae}(c)), we are able to manipulate the stiffness of the chain. 
When constrained at zero velocity, the microbot exerts a force in the direction of its polarisation which is parallel to the chain's axis at rest and point in the same direction, towards the anchor point of the chain (FIG.\ref{fig:ae}(d)). 

\textit{Transition to self-oscillations.} --- The chain with the largest width in between the active particles was slightly pushed off from the equilibrium and stayed at the same position without further significant movements as shown in FIG.\ref{fig:ae} (e). We then gradually reduced the width of the connections, at $W=4.25\pm 0.1$mm, the self-oscillation behaviour started to emerge (FIG.\ref{fig:ae} (f)) suggesting a competition between activity and elasticity: Active forces from active particles destabilise the elastic chain, which in turn, through deformation, re-orient the polarisation of the particle, ultimately leading to self-oscillation. The magnitude of the oscillations increases drastically (shown in FIG.\ref{fig:ae} (g)) with decreasing $W$ thanks to the competition between buckling and active force. 
This oscillatory dynamics can be quantified by \rev{the mean curvature} $\Theta(t):=\sum_{i=1}^{N-1}\theta_{i+1}(t)-\theta_i(t)=\theta_7 - \theta_1$ and the mean polarization $\Omega(t):=\frac{1}{N}\sum_{i=1}^{N}\theta_{i}(t)$, where $\theta_i(t)$ is the instantaneous orientation of particle $i$ w.r.t. the vertical axis (FIG.\ref{fig:be} (a,b)). While chains with large $W$ come to a standstill, softer chains exhibit a limit cycle (FIG.\ref{fig:be} (c)). The area of this limit cycle arises directly from a balance of energy injection with dissipation.


\textit{Active pendulums model.} --- What is the origin of such transition? Inspired by the Ziegler destabilisation paradox in structural mechanics~\cite{Bosi2016,Hagedorn1970,Bigoni2011,Ziegler1952} and the existing numerical models of active filaments~\cite{Farkas2002,Isele-Holder2015,Winkler2017,Sekimoto1995}, we construct a discrete model, where we boil the complexity of the elastic interactions down to three-body bending forces between the particles and the complexity of the vibration-induced dynamics to viscous overdamped dynamics. The discrete model is based on a chain of seven pendulums (shown as FIG.\ref{fig:be}(d)) with one end fixed. The pendulums have a length $\ell$ and are connected to their neighbors via a torsional spring of torsional stiffness $C$. Each pendulum $i$ is driven by a constant active force $\mathbf{F^a_i}=-F^a (\cos\theta_i \mathbf{e_x}+\sin\theta_i \mathbf{e_y})$ exerted on its end and in \rev{the direction of the pendulum}. We also introduce isotropic viscosity $\gamma$ contributing to a dissipative force on the end of each pendulum that is only dependent on its velocity, and assume no inertia in the system~{\footnote{A more accurate description would include additional effects such as noise in the active force and anisotropic viscous drag. In the former case, the noise is multiplicative in the equations of motion, which in turn progressively introduces noise to the trajectory of the active chain as the noise of the active force is increased, See SI.}}.
We then collected all the terms in $\delta\theta_i$ for each $i$ according to Virtual-Work Theorem and constructed $N$ nonlinear coupled DAEs that describe the motion of the elasto-active chain (further details in SI). 
There $\sigma=F^a\ell/C$ is the elasto-active parameter and $\tau=\gamma \ell^2/C$ a characteristic timescale. We estimate the torsional stiffness $C$ from their geometry using beam theory (See Appendix A). From the average velocity of the robots when they are freely moving $v_\textbf{a}=0.025\pm 0.005$m/s and their average force when they are pinned $F_\textbf{a}=15.7\pm3.1$mN (Fig. ~\ref{fig:ae}(d)), we estimate the damping coefficient $\gamma=F_\textbf{a}/v_\textbf{a}=0.63\pm 0.11$N.s/m. 

Using these parameters, we solve the system of DAEs numerically (See SI) and find a good agreement with the experimental results, and for the time series of average polarisation (Fig.~\ref{fig:be}(e)). We observe an agreement between the experiments and the simulation in the trend of the {limit cycle} (Fig.~\ref{fig:be}(f)), here the differences are due to the energy loss in the experimental scenarios. This agreement shows that nonlinear geometry, torsional stiffness, active force and isotropic viscous dissipation are sufficient ingredients to successfully capture the essence of the self-oscillation phenomenon. {The critical elasto-active number decreases with the length of the chain as $\sim 1/N^3$, See SI\footnote{See Supplemental Material [url] for simulations and theory, which includes 
Refs. \cite{young2002roark,Sekimoto1995,chelakkot2014flagellar,DeCanio_JRSI2017,HolmesBook,niedermayer2008synchronization}.} for simulations and theory, this can be interpreted by two effects: the Euler buckling load decreases for longer chains $\sim 1/N^2$ while the sum of the active forces grows as $N$.}
Our elasto-active model is controlled by a single timescale $\tau$ and a single non-dimensional parameter $\sigma$, which will allow us to probe the nature of the transition to self-oscillations in the following.

\begin{figure}
 \begin{center}
  \includegraphics[width=\linewidth,clip,trim=0cm 0cm 0cm 0cm]{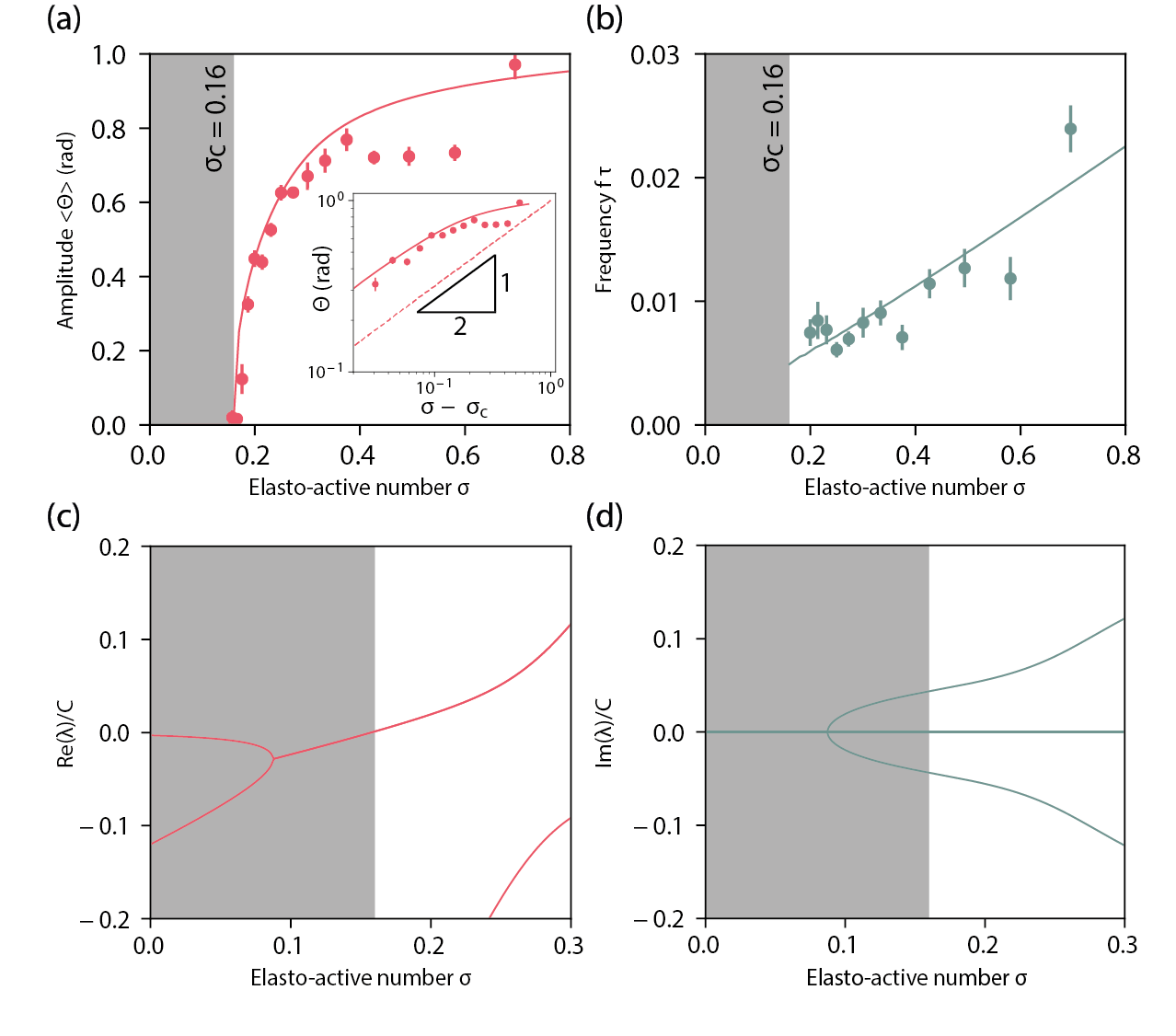}
 \end{center}
 \caption{An active system featuring a Hopf bifurcation at $\sigma=0.15$.
 Simulation and experimental results showing the evolution of (a) amplitude and (b)frequency with increasing $\sigma$ respectively for the chain with $N=7$. The inset in (a) is a log-log plot demonstrating the power law between $\Theta$ and $\sigma$. (c) Real and imaginary (d) part of the eigenvalues of the Jacobian of Eqs.~(\ref{eq:sysN2a})-(\ref{eq:sysN2b}) vs. elasto-active number $\sigma$ for a minimal chain with $N=2$. In all panels, the gray area represents the stable region. }
\label{fig:ab}
\end{figure}

\textit{Super-critical Hopf bifurcation.} --- We ran experiments and simulations over a wide range of the elasto-active parameter $\sigma$,
collected the time average of the amplitudes $\langle\Theta\rangle$ (Fig.~\ref{fig:ab} (a)) and the rescaled oscillation frequency $f \times\tau$ (Fig.~\ref{fig:ab} (b)) in the mean polarisation time series and plotted them against elasto-active parameter $\sigma$. While for low values of $\sigma$, the chain remains straight without oscillations
, we see that above a critical value $\sigma_c=0.16\pm 0.005$, the oscillation amplitude $\langle\Theta\rangle$ increases rapidly as $\langle\Theta\rangle\sim(\sigma-\sigma_c)^{0.5}$ (Fig.~\ref{fig:ab} (a-inset)), while the rescaled frequency increases linearly. To further elucidate the nature of the transition to self-oscillations, we carry out a linear stability analysis on the set of nonlinear coupled equations (See SI), and observe that at $\sigma=0.15$, the real part of a pair of eigenvalues becomes positive, while the corresponding imaginary parts of these eigenvalues are equal and opposite and monotonically increase (Fig.~\ref{fig:ab}(cd)). This transition is a hallmark of a Hopf-bifurcation. The exponent $0.5$ in the experimental and numerical data suggests that this bifurcation is supercritical. To verify the nature of the bifurcation theoretically, we restrict our attention to two pendulums with $N=2$, which is the simplest case where the model could exhibit the bifurcation. The time-evolution of such elasto-active chain is governed by the following equations 
\begin{eqnarray}
\tau\!\left(2 \dot{\theta _1}\! +\!\dot{\theta _2}  \cos \left(\theta _1\!-\!\theta _2\right)\right)&=&\theta _2\!-\!2 \theta _1+
\sigma  \sin \left(\theta _1-\theta _2\right)\label{eq:sysN2a}\\
\tau\!\left(\dot{\theta _1} \cos \left(\theta _1\!-\!\theta _2\right)\!+\!\dot{\theta _2} \right)&=&\theta _1\!-\!\theta _2\label{eq:sysN2b}.
\end{eqnarray}
In SI, we use a perturbative expansion and perform a few algebraic manipulations to demonstrate that Eqs.~(\ref{eq:sysN2a}-\ref{eq:sysN2b}) can be mapped onto the Landau-Stuart equation
\begin{equation}
\frac{dz}{dt}= (i+\sigma-3) z + \left(
i\left(\frac{17}{4}-\sigma\right)-\left(\sigma-\frac{5}{2}\right)\right) |z|^2 z,
\end{equation}
where $z$ is the complex variable defined by $z:=\theta_1+i\sqrt{(\sigma-2)/(4-\sigma)}\theta_2$. This equation is the canonical form of a supercritical Hopf-bifurcation. 
Many earlier works had observed experimentally or numerically self-oscillation phenomena~\cite{machin1958wave,lo2013mechanism,nishiguchi2018flagellar} or theoretically proposed models with Hopf-bifurcations~\cite{Sekimoto1995,camalet1999self,hilfinger2008chirality,hilfinger2009nonlinear,chelakkot2014flagellar,gonzalez2018active}; here, we unambiguously demonstrate experimentally, numerically and theoretically in a single system of active chains that the supercritical Hopf bifurcation underlies the transition to self-oscillations and is primarily controlled by the elasto-active number $\sigma$.

\begin{figure}[t!]
 \begin{center}
  \includegraphics[width=1.0\columnwidth,trim=0cm 0cm 0cm 0cm]{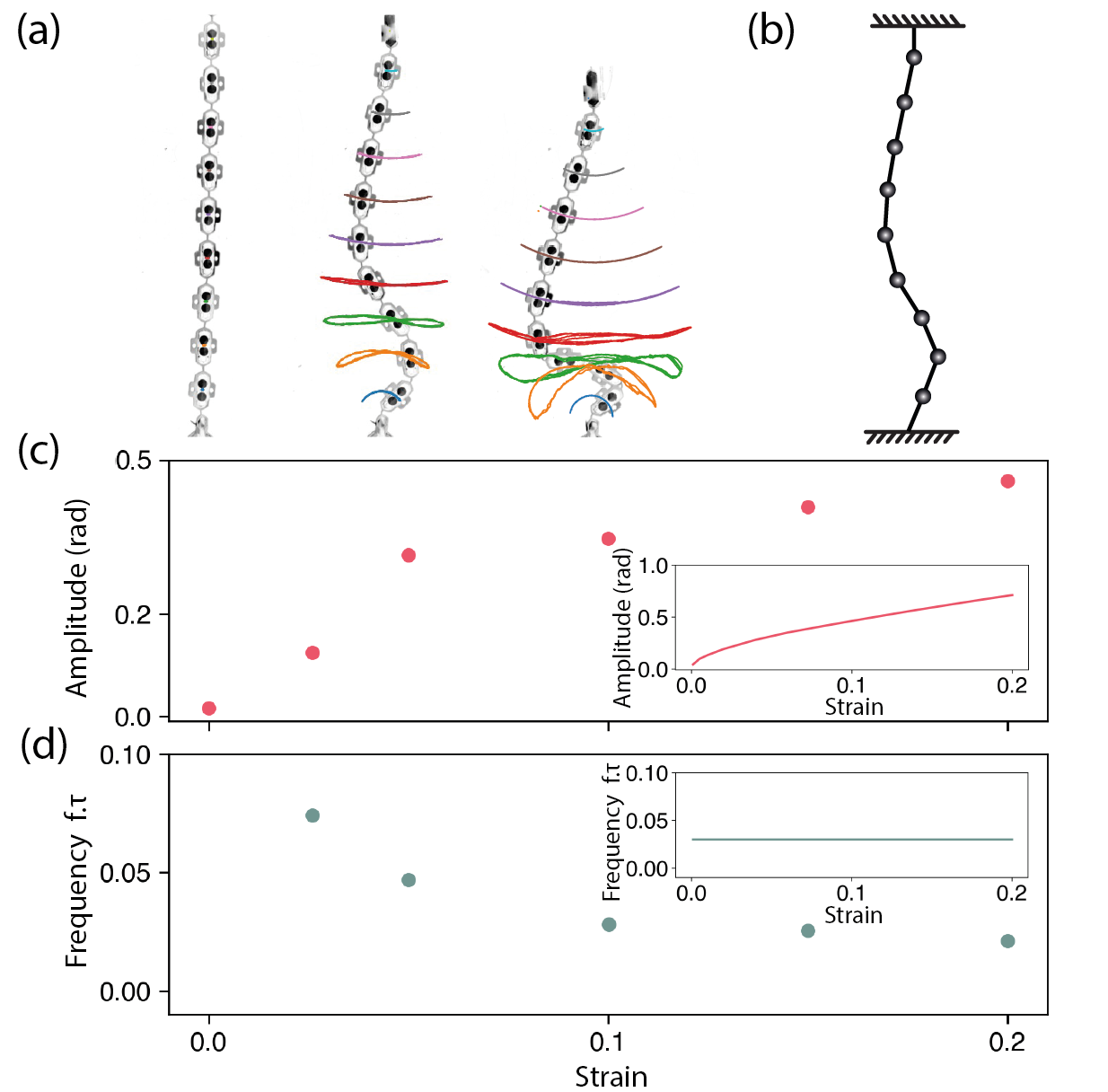}
 \end{center}
 \caption{\rev{{\bf Elasto-active chains pinned at both ends.} Stills of an active chain where both ends are pinned with overlaid trajectories of the hexbugs. The end-to-end distance between the pinning points is $\ell=500$ mm (left), $\ell=475$ mm (middle) and $\ell=400$ mm (right). The length of the undeformed chain is $L=500$ mm. (b) Model: sketch of the chain pinned at both ends. (c) Amplitude and (d) frequency of the self-snapping oscillations vs. the compressive strain $(L-\ell)/L$. The inset are the corresponding data from the numerical model defined in the SI. The elasto-active number of the chain is $\sigma=0.8$. See also Supplementary Videos.}}
\label{fig:figure5}
\end{figure}
\rev{\textit{Self-snapping.} --- When experimenting with the elasto-active chain, we realized that not only does it oscillate when pinned at one end, it also does oscillate when pinned at two ends, Fig.~\ref{fig:figure5}a. We find that these oscillations vanish when the chain is maintained at its undeformed length, but that they immediately emerge once we compress the chain along its axis. A passive chain would simply buckle, i.e. bend sideways when compressed. In stark contrast, the elasto-active chain bends sideways, but continuously snaps by itself from one side to the other. The more the chain is compressed, the more it oscillates (Fig.~\ref{fig:figure5}c) and the slower it self-snaps  (Fig.~\ref{fig:figure5}d). We find that adding geometrical constraints to our model (See SI) allows us to reproduce the phenomenology qualitatively (Fig.~\ref{fig:figure5}b)~\footnote{We find that the model overestimates the magnitude of oscillations by a factor 2 and that it fails to capture the decrease of self-snapping frequency as the chains is more compressed. We attribute the discrepancy between the experiment and the model to the fact we assumed the friction to be isotropic. In reality, the lateral drag is much larger than the longitudinal drag and this simplifying hypothesis can plausibly lead to predictions that overestimate the magnitude and the frequency of snapping.}.
While the flagellar motion observed earlier is ubiquitous in the context of biological structures, this self-snapping oscillation of buckled elasto-active structures is much more rare and surprising.}

\begin{figure}
 \begin{center}
  \includegraphics[width=.99\linewidth,clip,trim=0cm 0cm 0cm 0cm]{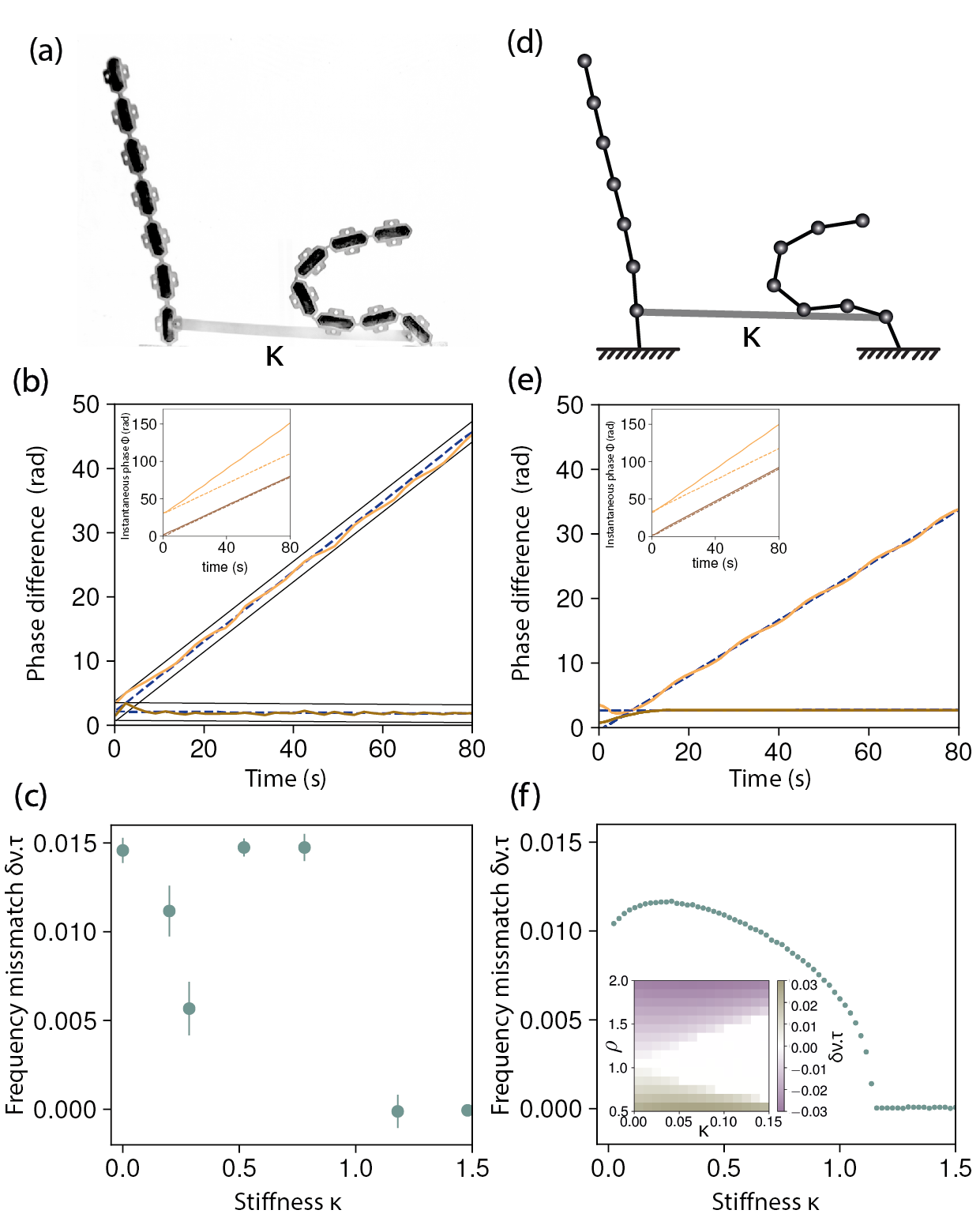}
 \end{center}
 \caption{{\bf Synchronisation of two elasto-active chains with different elasticity coupled by the first particles only.} (a) Snapshot of a pair of elasto-active chains with different elasticity coupled by another stiff silicon rubber chain. (b) {Evolution of the instantaneous phase differences (instantaneous phases $\Phi_1$ (dashed lines) and $\Phi_2$ (solid lines) in inset) of two elasto-active chains with coupling strength $K=0.8$ (yellow lines) and $K=1.2$ (brown lines)}. (c) Instantaneous frequency difference extracted from the instantaneous phase difference $\Psi$ vs. rescaled coupling stiffness $\kappa${, with error bars corresponding to deviations on the linear regression (dashed blue line and black lines in panel b)}. (d) Schematics of the numerical model adding a coupling spring ($K$) to two previously established elasto-active chains. (e)-(f) Same data as (b)-(c) for the numerical simulations. {An Arnold tongue is observed when the frequency miss-match is plotted against the ratio between both elasto-active numbers $\rho$ and the stiffness $\kappa$ (inset of panel f).}}
\label{fig:ke}
\end{figure}

\emph{Frequency entrainment synchronisation transition.} --- Our model system also allows us to explore synchronisation phenomena between two active chains (FIG.\ref{fig:ke} (a)-(d)). 
We demonstrate experimentally and numerically that an elastic coupling allows for a frequency entrainment synchronization transition. 
We selected two chains both with an elasto-active number $\sigma_1=0.8$ (the chain on the left hand side in FIG \ref{fig:ke} (a)) and an elasto-active number  $\sigma_2=0.6$ (the chain on the right  hand side in FIG \ref{fig:ke} (a)) and connected them via a coupling spring of variable stiffness $K$. We also performed simulations over a range of stiffness $K$ that contains what we have utilised in the experiments (FIG.\ref{fig:ke} (e)-(f)). The rescaled coupling stiffness is $\kappa:=K\ell^2/C_1$ (where $C_1$ is the stiffness of the left chain). To analyze the synchronization transition, we first extracted the oscillation signals from both chains. We then calculated the instantaneous phases $\Phi_1(t)$ and $\Phi_2(t)$ (see Appendix A) of each timeseries (Fig. \ref{fig:ke}(b) and (e) for experiments and simulations respectively). For low coupling stiffness (yellow lines), both $\Phi_1(t)$ and $\Phi_2(t)$ increase linearly, but with different slope, which can be described as two chains oscillating with different frequencies. On the contrary, for large coupling stiffness (brown lines), both instantaneous phases $\Phi_1(t)$ and $\Phi_2(t)$ align on the lowest slope (i.e. both chains beat at the lowest frequency of the two). We performed experiments and numerical simulations and measured the frequency mismatch $\delta\nu$ from the slope of the instantaneous phase difference $\Psi:=\Phi_2(t)-\Phi_1(t)$ over a wide range of coupling stiffness and found that the synchronization transition occurs at the  critical value $\kappa=1.1$ (Fig. \ref{fig:ke}(c) and (f)) \footnote{For the experimental results, the dip in the frequency miss-match at intermediate stiffness is due to friction effect that could not be avoided for small beam stiffness}. {Further numerical simulations of elasto-active chains with varying $\sigma$ reveal that the regime of synchronization exhibits an Arnold tongue centered about the 1:1 frequency ratio in the synchronization regime (Fig.~\ref{fig:ke} (f) inset). In other words, the two chains will synchronize for lower coupling if they have similar elasto-active numbers or if they are closer to the bifurcation---in this case the \rev{transition} towards steady oscillatory synchronized state will take longer.}

To rationalize this finding, we show in SI that the instantaneous phase difference $\Psi(t)$ between two \rev{chains} with $N=2$ is 
\begin{equation}
    \frac{d\Psi}{dt}=d\nu - \frac{\varepsilon}{\cos\Psi_0}\sin(\Psi-\Psi_0),\end{equation}
where $d\nu$, $\varepsilon$ and $\Psi_0$ are functions of the elasto-active number of each chain and of the coupling stiffness between the chains (See SI for closed forms). This equation has been well studied before for the investgation of synchronization phenomena~\cite{pikovsky2002synchronization}. An analysis of this equation predicts synchronization for $|\varepsilon/\cos\psi_0|>|d\nu|$, with a square root singularity~\cite{pikovsky2002synchronization}. As we show in the appendix, this condition is met when the coupling stiffness exceeds the threshold value $\kappa_c=11/(4\sqrt{6}) \sqrt{(\sigma-3) |1-\rho |}$,  
where $\sigma$ is the elasto-active number of the left chain and where $\rho$ is ratio between the elasto-active number of the right chain over that of the left chain. This result thus demonstrates that the synchronization scenario of the two elasto-active chains corresponds to that of a classic nonisochronous synchronization, which is characterized by a constant phase shift in the synchronized region. In addition, the two chains will synchronize for lower coupling if they are closer to the bifurcation or when they have similar elasto-active numbers. In summary, we have captured experimentally, numerically and analytically the synchronization transition of two elastically coupled active chains.

\textit{Conclusion.} --- 
In conclusion, we have shown that elasto-active chains exhibit transitions to self-oscillations and synchronization. 
{Since they exhibit a nonlinear dynamics that is governed by activity, elasticity and viscous damping, our study establishes macroscopic active structures as a powerful tool to investigate dynamical and autonomous behavior of active solids and living matter that exhibit collective self-oscillation at the microscale.}
Finally, fascinating future research directions could be taken from the minimalistic system considered here. For instance, one could investigate more complex geometries, such as non-follower forces~\cite{Baconnier_Arxiv2021}, more intricate geometrical connections between active particles, two-dimensional structures, alternative boundary conditions---pinned or moving clamping points (See Appendix A)---or even mechanical responses such as longitudinal or transverse excitations. \rev{These could in particular emulate peculiar dynamics observed in other contexts, such as odd elasticity \cite{Scheibner_NatPhys2020} or the non-Hermitian skin effect~\cite{Brandenbourger_NatComm2019,Coulais_NatPhys2021}.}

\textit{Acknowledgments.} --- We thank D. Giesen, S. Koot and all the staff members from the Technology Centre of University of Amsterdam for their technical support. We acknowledge K. Sekimoto,  A. Deblais, R. Sinaasappel, D. Tam  and  M. Jalaal  for insightful discussions. We acknowledge funding from the Netherlands Organization for  Scientific  Research through support from the NWO (Vidi grant no. 680-47-554/3259) and from the European Research Council Starting Grant No. ERC-StG-Coulais-852587-Extr3Me (C.C.). {The data and codes supporting this study are publicly available on Zenodo \rev{at https://doi.org/10.5281/zenodo.6549887.}}

\bibliography{biblio.bib}

 

\end{document}


\title{Self-oscillation and Synchronisation Transitions in Elasto-Active Structures: Supplementary Information}

\author{Ellen Zheng}
\affiliation{Institute of Physics, Universiteit van Amsterdam, Science Park 904, 1098 XH Amsterdam, The Netherlands}
\author{Martin Brandenbourger}
\affiliation{Institute of Physics, Universiteit van Amsterdam, Science Park 904, 1098 XH Amsterdam, The Netherlands}
\author{Louis Robinet}
\author{Peter Schall}
\affiliation{Institute of Physics, Universiteit van Amsterdam, Science Park 904, 1098 XH Amsterdam, The Netherlands}
\author{Edan Lerner}
\affiliation{Institute of Physics, Universiteit van Amsterdam, Science Park 904, 1098 XH Amsterdam, The Netherlands}
\author{Corentin Coulais}
\affiliation{Institute of Physics, Universiteit van Amsterdam, Science Park 904, 1098 XH Amsterdam, The Netherlands}
  
\maketitle

\tableofcontents

\clearpage
\setcounter{equation}{0}
\renewcommand{\theequation}{A\arabic{equation}}%
\setcounter{figure}{0}
\renewcommand{\thefigure}{A\arabic{figure}}%

\section{{Appendix A. Supplementary Videos}}

\begin{table}[h!]
{
\begin{tabular}{ccc}
Filename& Ligament width $W$ & Elasto-active number $\sigma$\\
\hline
Supplementary Video 1 & 5.0& 0.17\\
Supplementary Video 2 & 4.4& 0.21\\
Supplementary Video 3 & 2.0& 0.80\\
\end{tabular}
\caption{Supplementary Movies 1 to 3: 
Experimental Movies of a single elasto-active chain with 7 active particles and of varying thickness.}}
\end{table}
\\

\begin{table}[h!]
{
\begin{tabular}{cc}
Filename& Stiffness of connecting ligament $\kappa$ \\
\hline
Supplementary Video 4 & 0.20\\
Supplementary Video 5 & 1.48\\
\end{tabular}
\caption{Supplementary Movies 4 to 5: Experimental Movies of coupled elasto-active chains with 7 active particles each and two different connecting ligaments with different spring constant.}}
\end{table}

Supplementary Movies 6 and 7 corresponds to the pictures shown in \rev{Fig. A2a and Fig. A2b of the SI}, respectively. \rev{Supplementary Movies 8 to 10 correspond to the pictures shown in Fig 4a of the Main Text (from left to  right respectively).}

\section{Appendix B. Experimental methods}
\subsection{Construction of one-dimensional elasto-active chain}
Our elasto-active chain is simply constructed by 7 active units in a laser-cut silicon rubber chain. We fix one end of the chain with a clamp mechanically fixed by silicon glue on a 1.5mx1.5m black PET board. A camera (Basler acA2040-90$\mu$m) was mounted on an aluminium profile above the chain to track its motion.

\textit{active units} - HEXBUG nano\textsuperscript{\textregistered}
(random color) plays the role of active particle in our system. It is a self-propelled minirobot powered by a tiny motor (with AG13/LR44 1.5V button battery) and 12 rubber legs as shown in Fig. 1.(b). \david{We measured the pulling force of a single microbot on a tensile test machine (Instron 5940 Series, load cell 5N with a resolution of $0.5$mN) above a PET board, the results fluctuated rapidly due to the impulsive movement of the microbot and exhibited Gaussian distribution with a mean of $15.7$mN and a standard deviation of $3.1$mN  (shown as FIG.1 (d) of the Main text). }

\textit{Laser-cut rubber chain} - We first constructed the models of the rubber chains using Autodesk Inventor\textsuperscript{\textregistered} then printed the model with Universal Laser PLS6.150D. Width of the connection in the model increased from 2mm to 5mm with a 0.2mm step, the rest of the geometries unchanged. The 'ears' on the side of the constraining units were designed for the convenience of connecting two single chains together. The circular end was made to ease the fixation to the clamp. Real geometries of the chains printed from the laser cutter differs slightly with the geometries in the model, they were measured again using vernier caliper. The flexibility of the silicone rubber chain brought negligible errors in geometry measurements, the real thickness was set to be the value at which the chain can not be clipped by the caliper anymore. {We clamped the head of the active chain to a metallic bar. This clamping effectively connects the first active particle to the laboratory frame with a flexible hinge of stiffness $C$. Upon bending, the chain remains in the plane with no significant warping while bending and oscillating.}

\subsection{Realisation of the coupled elasto-active chains}
Taking two elasto-active chains with W = 2mm and 2.8mm respectively, we connected the two of them with different coupling chains. The coupling chain was fixed onto the 'ears'of the elasto-active chains with plastic pins, the elasto-active chains themselves were fixed by the clamps in the same way as the singular chains. The highest stiffness was provided by a simple rubber chain (380mmx15mm), we then tuned the stiffness by adding triangular teeth to the simple geometry. The coupling chain with 1 teeth possessed the lowest stiffness while adding more teeth to it slightly increased the stiffness. Schematics of the chains can be found in the supplementary materials. 

\subsection{Calibration and measurements}
\textit{Torsional Stiffness} - As mentioned in the main paragraph, we measured the Young's modulus of the silicon rubber with Instron 5940 Series at a strain rate of 0.05 mm/min. Torsional stiffness (C) of the connections shown in FIG.\ref{fig:ae}(c) was calculated \cite{young2002roark} with:
\begin{equation}
C=\frac{GJ}{L}
\end{equation}
where G was taken as the Young's modulus (0.239MPa) and J is the torsional constant that was determined with:
\begin{equation}
J=w^3h(\frac{16}{3}-3.36\frac{2}{h}(1-\frac{w^4}{12h^4}))
\end{equation}
where h and w was half the value of H and W shown in FIG.1 of the main text.
\\
\textit{Tracking of Motion} - Motion of the active chains was recorded by the camera (Basler acA2040-90$\mu$m) at a frame rate of 60fps and resolution of $4$Mpx. The images were binarized and eroded, we then detected and tracked the active units using the opencv module under Python.

\textit{Instantaneous phase} - \rev{The exact oscillation frequency of the two coupled active-chains was obtained by computing their  instantaneous phase over time. For each chain, we expressed the oscillation amplitude, described here by the mean curvature  $\Theta(t)$ in its analytical form $\Theta_{\mathrm {a} }(t)$: 
\begin{equation}
\Theta_{\mathrm {a} }(t)=\Theta(t)+j{\hat {\Theta}}(t)=\Theta_{\mathrm {m} }(t)e^{j\phi (t)},
\end{equation}
where ${\displaystyle {\hat {\Theta}}(t)\triangleq \operatorname {\mathcal {H}} [\Theta(t)]}$ is the Hilbert transform of the signal and $\Theta_{\mathrm {m} }(t)$ is the instantaneous amplitude of the envelope. The instantaneous phase  of the signal corresponds to ${\displaystyle \phi (t)\triangleq \arg \!\left[\Theta_{\mathrm {a} }(t)\right]}$ and is plotted in Fig. 4b\&c in the main text. The instantaneous frequency was calculated via a linear regression on the instantaneous phase. The Fig 4c\&e show the mismatch between the two instantaneous frequencies of the chains.}

\rev{
\subsection{Additional experiments}
We performed complementary experimental observations with our chains.
\subsubsection{Chains pinned at one end with $N=17$.}
We first investigated the dynamics of longer chains with an elasto-active number $\sigma=0.8$, Fig.~\ref{fig:longerchain}. We see that a similar oscillatory motion occurring, yet in addition to the lowest harmonic seen in the short chain and in the stability analysis, higher harmonic seem to be occurring, which we speculate are due to self-contact interactions.
\begin{figure}[h!]
 \begin{center}
  \includegraphics[width=.66\linewidth,clip,trim=0cm 0cm 0cm 0cm]{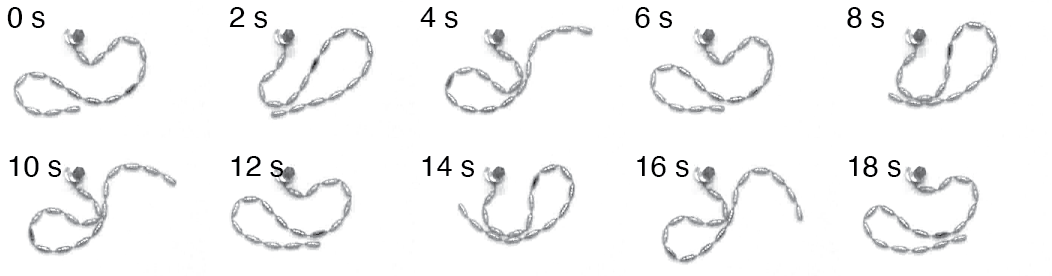}
 \end{center}
 \caption{\rev{{{\bf Self-oscillations in an elasto-active chain with $N=17$.} We observe a similar pattern of oscillations as in the case $N=7$ described in the Main Text.}}}
\label{fig:longerchain}
\end{figure}
}
\rev{
\subsubsection{Chains confined on a freely rotating pin and in a paraboloid.}
We also explored the role of various confinement by pinning elasto-active chains to a freely rotating clamp, Fig~\ref{fig:confined}a and in a paraboloid, Fig.~\ref{fig:confined}b. We observe that, when the chain is pinned to a freely rotating clamp, it rotates at a constant speed around the pinning point, as was predicted in~\cite{Sekimoto1995,chelakkot2014flagellar}. It picks up a spontaneous rotation orientation. Similarly, when the chain is confined in a paraboloid, it tends to rotate around the center of the paraboloid, but this time is also craws in circles at the same time. 
See also Supplementary videos 6 and 7.
\begin{figure}[h!]
 \begin{center}
  \includegraphics[width=.66\linewidth,clip,trim=0cm 0cm 0cm 0cm]{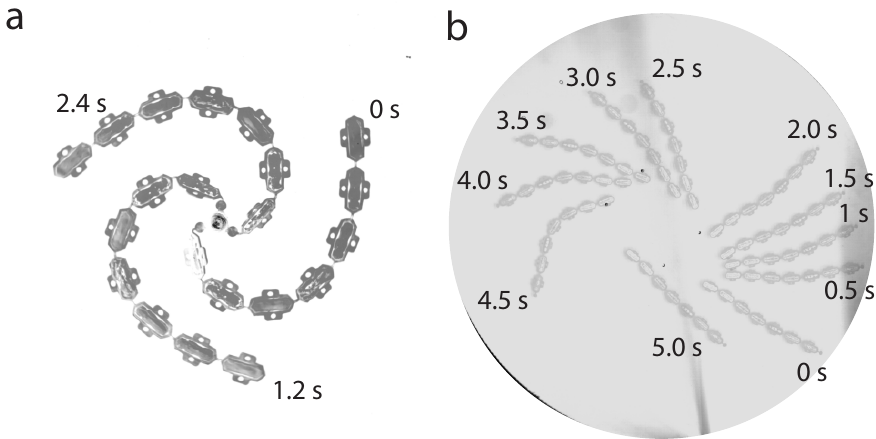}
 \end{center}
 \caption{\rev{{\bf Elasto-active chains in various confinements.}
Stills of an active chain (a) whose head is pinned to ball-bearing
($\sigma = 0.33$) and (b) in a paraboloid-shaped satellite
dish ($\sigma = 0.30$). In (a), the chain spins around the pinning
point and in (b) the chain spins around the paraboloid.}}
\label{fig:confined}
\end{figure}
}

\section{Appendix C.Theoretical Models}
\subsection{Derivation of equations of motion}

\subsubsection{Single elasto-active chain}
We aim to describe the active chain shown in Fig. 1a by a model of a elasto-active chain of $N$ pendula, with follower forces. The potential energy is
\begin{equation}
U=\frac{C}{2}\theta_1^2+\frac{C}{2}\sum_{i=1}^{N-1}(\theta_i-\theta_{i+1})^2
,\label{eq:PotentialEnergy}
\end{equation}
where $C$ is the torsional stiffness of the links between each pendulum and $\theta_i$ the angle of pendulum $i$ with respect to the $\mathbf{e_x}$ axis in Fig.~\ref{fig:ae}(d). Upon an infinitesimal variation of the internal degrees of freedoms, the change in potential energy is $\delta U=\sum_{i=1}^N (\partial U/\partial \theta_i) \delta\theta_i$ that yields
\begin{equation}
\delta U= C(2\theta_1-\theta_2)\delta\theta_1+C(2\theta_2-\theta_1\-\theta_3)\delta\theta_2+\cdots+C(2\theta_{N-1}-\theta_{N-2}\-\theta_{N})\delta\theta_{N-1}+C(\theta_N-\theta_{N-1})\delta\theta_{N}
.\label{eq:deltaPotentialEnergy}
\end{equation}
In addition, each particle $i$ located at the endpoint of each pendulum undergoes an active force $\mathbf{F}^a_i=- F^a (\cos\theta_i \mathbf{e_x} + \sin\theta_i \mathbf{e_y}) $, aligned with the pendulum $i$ and a dissipative force, which we assume is viscous drag $\mathbf{F}^d_i-\gamma (\dot{x}_i\mathbf{e_x} + \dot{y}_i\mathbf{e_y})$. $\gamma$ is the damping coefficient, $F$ the follower force exerted on each pendulum, $\ell$ the length of each pendulum and $x_i$ ($y_i$) the horizontal (vertical) displacement of the end point of pendulum $i$. $\dot{}$ denote the time derivative. Therefore, the work of these non-conservative forces upon infinitesimal variation of the internal degrees of freedoms $\{\delta x_1,\delta x_2,\cdots,\delta x_N, \delta y_1,\delta y_2,\cdots,\delta y_N\}$ of the system reads
\begin{equation}
\delta W=-\gamma \sum_{i=1}^{N}
\left(\dot{x}_i \delta x_i +
\dot{y}_i \delta y_i\right) 
-F^a\sum_{i=1}^{N}
\left(\cos\theta_i \delta x_i + 
\sin\theta_i \delta y_i \right),
\label{eq:IncWork}
\end{equation}
 Thanks to the geometrical relations $x_i= \ell \sum_{j=1}^i \cos\theta_j$ and $y_i= \ell \sum_{j=1}^i \sin\theta_j$,
 we can substitute
$\delta x_i= \ell \sum_{j=1}^i -\sin\theta_j\delta\theta_j$, 
$\delta y_i= \ell \sum_{j=1}^i \cos\theta_j\delta\theta_j$, 
$\dot{x}_i= \ell \sum_{j=1}^i -\sin\theta_j\dot{\theta}_j$, 
$\dot{y}_i= \ell \sum_{j=1}^i \cos\theta_j\dot{\theta}_j$, and Eq.~(\ref{eq:IncWork}) can be rewritten as a function of the angles
\begin{equation}
\delta W=-\gamma \ell^2\sum_{i=1}^{N}\sum_{j=1}^{i}\sum_{k=1}^{i}
\left(\cos(\theta_j-\theta_k)\dot{\theta}_j\delta\theta_k\right) 
-F^a\ell\sum_{i=1}^{N}\sum_{j=1}^{i} \left(\sin(\theta_i-\theta_j)\delta\theta_j \right).
\label{eq:IncWork_1}
\end{equation}
According to the Virtual-Work Theorem, at mechanical equilibrium, the work of external forces $\delta W$ (Eq.~(\ref{eq:IncWork_1})) is equal to the change of potential energy $\delta U$ (Eq.~(\ref{eq:deltaPotentialEnergy})). Collecting all the terms in $\delta\theta_i$ for each $i$,  With dimensionless parameters: $\sigma = F^a\ell/C$ and $\tau = \gamma l^2/C$, we find $N$ nonlinear coupled ordinary differential equations that describe the motion of the elasto-active chain.
\begin{eqnarray}
0&=&2\theta_1\!-\!\theta_2 \!-\! \sigma\sum_{j=1}^N\!\!\sin(\theta_1\!-\!\theta_j) \!+\! \tau  \!\left(\!N\dot{\theta}_1+\sum_{j=2}^N \dot{\theta}_j\!\cos(\theta_1\!-\!\theta_j)\!\right)\textrm{for }i\!=1\label{eq:motion_1}\\
\nonumber 0&=& 2\theta_i\!-\!\theta_{i-1}\!-\!\theta_{i+1} \!-\! \sigma\sum_{j=i}^N\sin(\theta_i\!-\!\theta_j)\\
&&+\tau \!\left(\!\!(\!N\!-\!i\!+\!1\!)\!\sum_{j=1}^i\!\dot{\theta}_j\!\cos(\theta_i\!-\!\theta_j)\!+\!\sum_{j=i+1}^N \!(\!N\!-\!j\!+\!1\!)\!\dot{\theta}_j\!\cos(\theta_i\!-\!\theta_j)\!\right) \textrm{for }i\!\in\![2,N-1]\label{eq:motion_i}\\
0&=& \theta_N\!-\!\theta_{N-1} \!+\! \tau\sum_{j=i}^N \!\dot{\theta}_j\!\cos\left(\theta_N\!-\!\theta_j\right) \textrm{for }i\!=N,\label{eq:motion_N}
\end{eqnarray}
which are the equations that we solve numerically in Figs. 2 and 3 or the Main Text.  \rev{Note that this is a system of differential algebraic equations (DAE), which we we solve using Mathematica.}

\subsubsection{Coupled elasto-active chains}
For a pair of coupled elasto-active chains with different elasticity, we took the work done by the dissipative force $F_d$ and the active force $F_a$ as a sum of these values of both chains. The resulting $\delta W$ is thus:\\
\begin{eqnarray}
\nonumber\delta W = &-&\gamma \ell^2 \left(\sum _{i=1}^{N} \left(\sum _{j=1}^i \left(\sum _{k=1}^i \delta \theta _k \dot\theta _j \cos \left(\theta _j-\theta _k\right)\right)\right)\right)+\sum _{i=1}^{N} \left(\sum _{j=1}^i \left(\sum _{k=1}^i \delta \phi _k \dot\phi _j \cos \left(\phi _j-\phi _k\right)\right)\right)\\
&-& F^a\ell \left(\sum _{i=1}^{N} \left(\sum _{j=1}^i \delta \theta _j \sin \left(\theta _i-\theta _j\right)\right)+\sum _{i=1}^{N} \left(\sum _{j=1}^i \delta \phi _j \sin \left(\phi _i-\phi _j\right)\right)\right)\label{eq:W_double},
\end{eqnarray}
where $\theta_i$ and $\phi_i$ depict the angle between each particle and the horizontal axis of individual chains. We then added the effect of the coupling force to the sum of $\delta U$ of both chains and rendered:
\begin{eqnarray}
\nonumber\delta U = &\theta _1&\left(\delta \theta _1 (C_1+K)-\delta \phi _1 K\right)+\phi _1  \left(\delta \phi _1 (C_2+K)-\delta \theta _1 K\right)+ C_1 \sum _{i=2}^{N} \delta \theta _i \left(\theta _i -\theta _{i-1} \right)+C_1 \sum _{i=1}^{N-1} \delta \theta _i \left(\theta _i -\theta _{i+1} \right)\\
&+& C_2\sum _{i=2}^{N} \delta \phi _i \left(\phi _i -\phi _{i-1} \right)+C_2\sum _{i=1}^{N-1} \delta \phi _i \left(\phi _i -\phi _{i+1} \right)\label{eq:U_double},
\end{eqnarray}
where $C_1$ and $C_2$ are the torsional spring constant in each chain and $K$ is the stiffness of the coupling chain. We introduced the following dimensionless parameters
\begin{eqnarray}
\rho = \frac{C_2}{C_1}\textrm{, }\kappa = \frac{K\ell^2}{C_1}\textrm{, }\sigma = \frac{F^a\ell}{C_1} \textrm{ and } \tau = \frac{\gamma \ell^2}{C_1},
\end{eqnarray}
which allowed to express the coupled ordinary differential equations as follows
\begin{eqnarray}
\nonumber 0 &=& \left(2+\kappa \right)\theta_1-\theta_2-\kappa\phi_1-\sigma\sum_{j=1}^N\!\!\sin(\theta_1\!-\!\theta_j)+\tau\left(N\dot\theta_1+\sum_{j=2}^{N}(N-j+1))\dot\theta_j cos(\theta_1-\theta_j)\right) \textrm{; }\\
\nonumber 0 &=& \left(2\rho+\kappa\right)\phi_1-\rho\phi_2-\kappa\theta_1-\sigma\sum_{j=1}^N\!\!\sin(\phi_1\!-\!\phi_j)+\\
&& \tau\left(N\dot\phi_1+\sum_{j=2}^{N}(N-j+1))\dot\phi_j cos(\phi_1-\phi_j)\right) \textrm{for }i\!=1\label{eq:motion_d1}\\
\nonumber 0 &=& 2\theta_{i} -\theta_{i-1}-\theta_{i+1}-\sigma_1\sum_{j=i}^N\!\!\sin(\theta_i\!-\!\theta_j)+\\
\nonumber && \tau\left(\!\!(\!N\!-\!i\!+\!1\!)\!\sum_{j=1}^i\!\dot{\theta}_j\!\cos(\theta_i\!-\!\theta_j)\!+\!\sum_{j=i+1}^N \!(\!N\!-\!j\!+\!1\!)\!\dot{\theta}_j\!\cos(\theta_i\!-\!\theta_j)\!\right) \textrm{; }\\
\nonumber 0 &=& \rho(2\phi_{i} -\phi_{i-1}-\phi_{i+1})-\sigma\sum_{j=i}^N\!\!\sin(\phi_i\!-\!\phi_j)+\\
&& \tau\left(\!\!(\!N\!-\!i\!+\!1\!)\!\sum_{j=1}^i\!\dot{\phi}_j\!\cos(\phi_i\!-\!\phi_j)\!+\!\sum_{j=i+1}^N \!(\!N\!-\!j\!+\!1\!)\!\dot {\phi}_j\!\cos(\phi_i\!-\!\phi_j)\!\right) \textrm{for  }i\!\in\![2,N-1]\label{eq:motion_di}\\
\nonumber 0&=& \theta_N\!-\!\theta_{N-1} \!+\! \tau\sum_{j=i}^N \!\dot{\theta}_j\!\cos\left(\theta_N\!-\!\theta_j\right) \textrm{;}\\
0&=& \rho(\phi_N\!-\!\phi_{N-1}) \!+\! \tau\sum_{j=i}^N \!\dot{\phi}_j\!\cos\left(\phi_N\!-\!\phi_j\right) \textrm{for }i\!=N.\label{eq:motion_d1N}
\end{eqnarray}
We solved again numerically \rev{this system of differential algebraic equations using Mathematica} in Fig. 5 of the Main Text. Below in the Section ``Synchronization of two coupled elasto-active chains", we considered the case $N=2$ and performed a perturbative expansion to map these ODEs onto an equation that describes the time evolution of the instantaneous phase difference between two elasto-active chains.

\subsection{Linear limit}
Eqs.~(\ref{eq:motion_1})-(\ref{eq:motion_N}) are impossible to solve explicitly in their full generality, but we can obtain some information about the behavior of the system by considering its linearisation close to the limit where all the angle are zero. 
\begin{eqnarray}
0&=& 2\theta_1\!-\!\theta_2 \!-\! \frac{Fa^2}{C}\sum_{j=1}^N\!\!(\theta_1\!-\!\theta_j) \!+\! \frac{\gamma a^2}{C} \!\left(\!N\dot{\theta}_1+\sum_{j=2}^N \dot{\theta}_j\!\right)\label{eq:motion_1_lin}\\
0&=& 2\theta_i\!-\!\theta_{i-1}\!-\!\theta_{i+1} \!-\! \frac{Fa^2}{C}\sum_{j=i}^N(\theta_i\!-\!\theta_j)\!+\! \frac{\gamma a^2}{C} \!\left(\!\!(\!N\!-\!i\!+\!1\!)\!\sum_{j=1}^i\!\dot{\theta}_j\!\!+\!\sum_{j=i+1}^N \!(\!N\!-\!j\!+\!1\!)\!\dot{\theta}_j\!\right)\textrm{for }i\!\in\![2,N-1] \label{eq:motion_i_lin}\\
0&=& \theta_N\!-\!\theta_{N-1} \!+\! \frac{\gamma a^2}{C} \sum_{j=i}^N \!\dot{\theta}_j.\label{eq:motion_N_lin}
\end{eqnarray}
The stability of the system can be investigated by injecting the following ansatz $(\theta_1,\theta_2,\dots,\theta_N)=(\theta^0_1,\theta^0_2,\dots,\theta^0_N)\exp\lambda t$
in Eqs.~(\ref{eq:motion_1_lin})-(\ref{eq:motion_N_lin}) and requiring that the determinant of such system of equation to be zero, which which can be rewritten in the following matrix form:
{$A.(\theta^0_1,\theta^0_2,\dots,\theta^0_N)^T=0$, with 
\begin{equation}
A_{ij}=(2\delta_{ij}-\delta_{i,j-1}-\delta_{i-1,j})+ \left((N-j+1) u_{ij} + (N-i+1) d_{ij}\right)\tau\lambda +\left(u_{ij}-(N-i) \delta_{ij}\right)\sigma,
\end{equation}
where $\delta_{ij}$ is the Kronecker delta, $u_{ij}$ the strictly upper triangular matrix and $d_{ij}$ the lower triangular matrix. By solving det$A=0$, we find the eigenvalues $\lambda$ as a function of the elasto-active number $\sigma$, see Fig. 3cd of the Main Text for the case $N=7$. The Hopf bifurcation occurs at $\sigma_c$, when the real part of $\lambda$ becomes positive. We plot below in Fig.~\ref{fig:sizeeffect} $\sigma_c$ vs. $N$ and find that $\sigma_c\sim 1/N^3$.}
{
\begin{figure}[h!]
 \begin{center}
  \includegraphics[width=.66\linewidth,clip,trim=0cm 0cm 0cm 0cm]{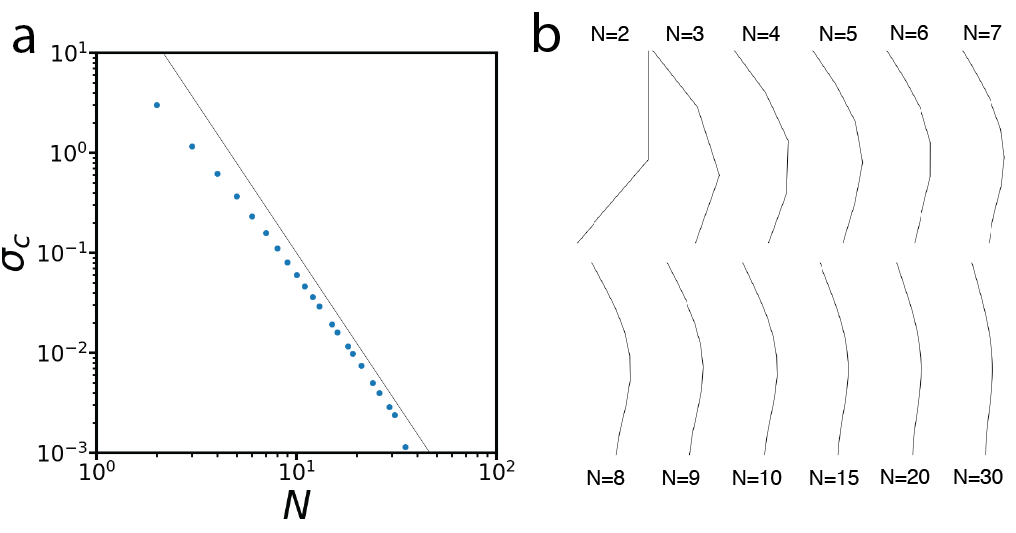}
 \end{center}
 \caption{{{\bf Size effects.} Critical elasto-active number $\sigma_c$---where the onset of the Hopf bifurcation occurs---vs. $N$ the number of active particles the chain is made of (blue markers). The black solid line indicate $\sigma_c\sim  N^{-3}$.}}
\label{fig:sizeeffect}
\end{figure}
}

{\subsection{Continuum limit}
We derive below the continuum equations in the linear limit. To derive the continuum limit of \MB{Eq.~(\ref{eq:motion_i_lin})}, we take $\theta_{i}=\theta(s)$ and $\theta_{i\pm 1}=\theta(s)\pm \MB{\partial} \theta \MB{(s)}/ \MB{\partial} s+1/2\,\MB{\partial}^2\theta\MB{(s)}/\MB{\partial}s^$. 
We obtain the following equations
\begin{equation}
0= -\frac{\partial^2\theta}{\partial s^2} -\sigma(\ell-s)\theta+\sigma\int_s^\ell ds' \theta(s') 
+ \tau\left(\int_0^s ds' (\ell-s)\frac{\partial\theta}{\partial t}(s') 
+\int_s^\ell ds' (\ell-s')\frac{\partial\theta}{\partial t}(s') 
\right)
\label{eq:eq_continuum}
\end{equation}
We then differentiate w.r.t to $s$ and use the chain rule:
\begin{equation}
0= -\frac{\partial^3\theta}{\partial s^3} +\sigma(\ell-s)\frac{\partial\theta}{\partial s} - \tau\left(\int_0^s ds' \frac{\partial\theta}{\partial t}(s') 
\right)
\label{eq:eq_continuumd}
\end{equation}
We then introduce the deflection of the chain from its rest position $v$ which is related to the local angle of the chain as  $\theta=\partial v/\partial s$. The above equations then become
\begin{equation}
0= -\frac{\partial^4 v}{\partial s^4} +\sigma(\ell-s)\frac{\partial^2 v}{\partial s^2} - \tau \frac{\partial v }{\partial t}.
\label{eq:eq_continuumdd}
\end{equation}
Eq.~(\ref{eq:eq_continuumdd}), together with the boundary conditions $v(0)=\partial v /\partial s(0)=\partial^2 v /\partial s^2(\ell)=\partial^3 v /\partial s^3(\ell)=0$, which correspond to a clamped head and a free tail, has been solved by Sekimoto et al.~\cite{Sekimoto1995} and revisited by Chekkalot et al.~\cite{chelakkot2014flagellar}. They both predict a Hopf bifurcation at a critical value of $\sigma_c$ that scales with $N^{-3}$, which is consistent with our data.}

{
\subsection{Role of Noise}
In Fig. 1d, we see that the active force has some significant noise. In this section, we verify what the role of noise on the self-oscillation dynamics is. To this end, we consider the $N=2$ case with a noise in the active force. The equations of motion then become
\begin{eqnarray}
\tau\!\left(2 \dot{\theta _1}\! +\!\dot{\theta _2}  \cos \left(\theta _1\!-\!\theta _2\right)\right)&=&\theta _2\!-\!2 \theta _1+
\sigma(1 +\eta(t)) \sin \left(\theta _1-\theta _2\right)\label{eq:sysN2a}\\
\tau\!\left(\dot{\theta _1} \cos \left(\theta _1\!-\!\theta _2\right)\!+\!\dot{\theta _2} \right)&=&\theta _1\!-\!\theta _2\label{eq:sysN2b},
\end{eqnarray}
where $\eta(t)$ is a Gaussian white noise of amplitude $a $. Below in Fig.~\ref{fig:noise} we plot the trajectory of the elasto-active chain for increasing values of $a$ and see that they increasingly become chaotic.}
{
\begin{figure}[h!]
 \begin{center}
  \includegraphics[width=1\linewidth,clip,trim=0cm 0cm 0cm 0cm]{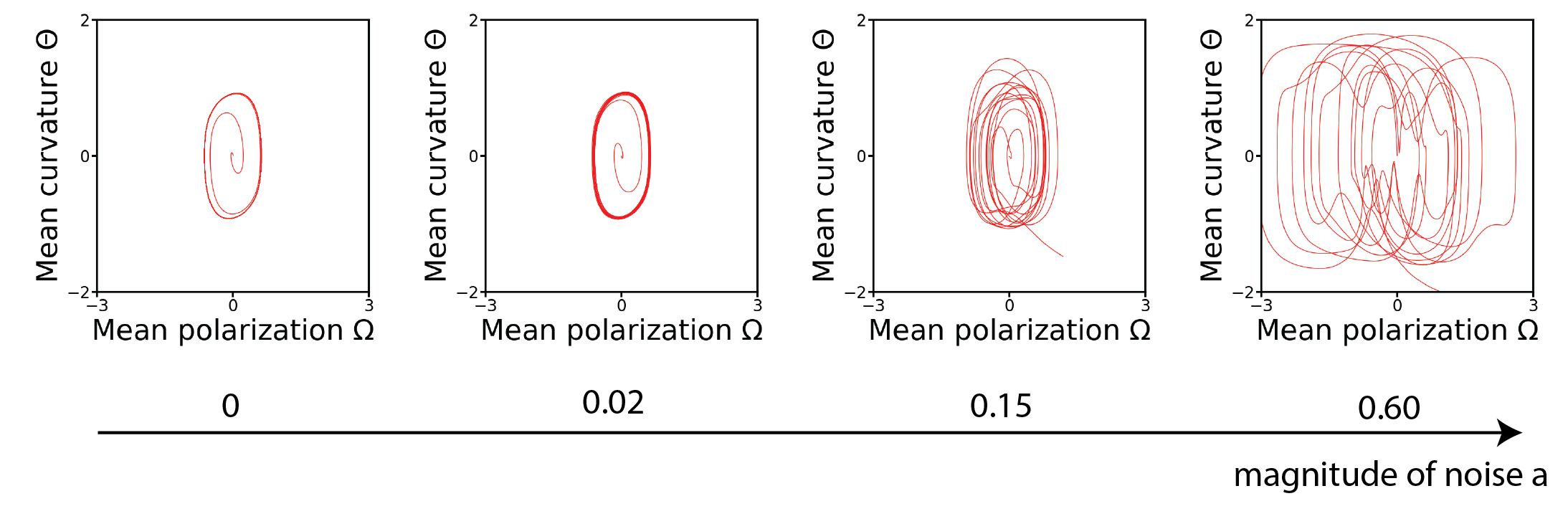}
 \end{center}
 \caption{{{\bf Role of noise.} Curvature $\Theta=\theta_2-\theta_1$ vs. mean polarization $\Theta=(\theta_1+\theta_2)/2$ of two coupled pendulums with follower forces and multiplicative noise a.}}
\label{fig:noise}
\end{figure}
}

{
\subsection{Role of anisotropic drag}
In this section, we derive the equations of motion of the elasto-active chain in the case where the drag is anisotropic. The assumptions are the same as in the section ``Derivation of equations of motion'' above, save for the fact that the drag force on particle $i$ is now of the form $\mathbf{F}^d_i=-\gamma^L \dot{v}^L_{i}\mathbf{e^L_{i}} - \gamma^T \dot{v}^T_{i}\mathbf{e^T_{i}}$, where $\dot{v}^L_{i}$ ($\dot{v}^T_{i}$) is the longitudinal (transverse) velocity along the axis of pendulum $i$ and $\gamma^L$ ($\gamma^T$) the corresponding drag coefficient. Cast in Cartesian coordinates, this drag force reads
\begin{equation}
   \mathbf{F}^d_i= \gamma^L
   \left(
   \begin{array}{c}
   \cos^2\theta_i\dot{x_i}+\cos\theta_i\sin\theta_i\dot{y_i}\\
   \cos\theta_i\sin\theta_i\dot{x_i}+\sin^2\theta_i\dot{y_i}
   \end{array}
   \right)+
   \gamma^T
   \left(
   \begin{array}{c}
   \sin^2\theta_i\dot{x_i}-\cos\theta_i\sin\theta_i\dot{y_i}\\
   -\cos\theta_i\sin\theta_i\dot{x_i}+\cos^2\theta_i\dot{y_i}
   \end{array}
   \right).
\end{equation}
Following the same steps as above, the work of the non-conservative forces becomes
\begin{equation}
\begin{split}
\delta W=&-\gamma^L\ell^2\sum_{i=1}^{N}\sum_{j=1}^{i}\sum_{k=1}^{i}
\left(\cos\theta_k \left(\sin ^2\theta _i \cos \theta _j-\sin \theta _i \cos\theta _i \sin \theta _j\right)
-\sin\theta _k \left(\sin \theta _i\cos \theta _i \cos \theta _j-\cos ^2\theta _i \sin\theta _j\right)\right)
\dot{\theta}_j\delta\theta_k
\\
&
-\gamma^T\ell^2\sum_{i=1}^{N}\sum_{j=1}^{i}\sum_{k=1}^{i}
\left(\cos \theta _k \left(\cos ^2\theta _i \cos \theta _j-\sin \theta _i\cos\theta _i \sin \theta _j\right)+\sin \theta _k \left(\sin ^2\theta _i \sin \theta _j+\sin \theta _i\cos\theta _i\cos \theta _j\right)\right)
\dot{\theta}_j\delta\theta_k\right)
\\
&
-F^a\ell\sum_{i=1}^{N}\sum_{j=1}^{i} \left(\sin(\theta_i-\theta_j)\delta\theta_j \right), \label{eq:IncWork_anis}
\end{split}
\end{equation}
and the dimensionless equations of motion then become in the case $N=2$
\begin{eqnarray}
\begin{split}
&\theta _1'\left(\!\tau^L \sin ^2(\theta _1\!-\!\theta _2)\!+\!\frac{\tau^T}{4} \left(\cos4 \theta_1+2 \cos 2 \theta _1 \cos 2 \theta _2\!+\!5\right)\!\right)
\\
&
\!+\!
\theta _2' \frac{\tau^T}{2} \left(2 \sin\theta _1 \sin \theta _2+\cos \theta _1 \left(\cos \theta _2+\cos 3 \theta _2\right)\right)
\end{split}
&=&\theta _2\!-\!2 \theta _1+
\sigma\sin \left(\theta _1-\theta _2\right)\label{eq:sysN2a}\\
\frac{\tau^T}{4}\!\left(
\left.\theta _2'\left(\cos 4\theta _2+3\right)+2 \theta _1' \left(\sin \theta _1 \left(\sin \theta _2-\sin 3 \theta _2\right)+2 \cos \theta _1 \cos \theta _2\right)\right)
\right)&=&\theta _1\!-\!\theta _2\label{eq:anisN2},
\end{eqnarray}
where $\tau^L=\gamma^L \ell^2/C$ and $\tau^T=\gamma^T \ell^2/C$. Interestingly, in the linear limit, we recover the same equation as for the isotropic drag case (Eqs. (1-2) of the Main Text). In other words, the transverse is the only one that matters for the onset of the Hopf bifurcation. However, the magnitude of the self-oscillations depends vastly on the anisotropy between the two directions of drags, see Fig.~\ref{fig:anisotropy} below.}
\rev{That the longitudinal drag does not affect the onset of instability can be interpreted by the fact the destabilization of the structure around its equilibrium position does not only entail longitudinal motion but only transverse motion. In contrast, that the magnitude of oscillations decreases with increasing longitudinal drag can be interpreted by the fact that oscillations of finite amplitude do involve both transverse and longitudinal motion. Our results are consistent with that of De Canio et al.~\cite{DeCanio_JRSI2017}, even though they use a different model where the follower forces are localized at the tip of the chain, while our follower forces are distributed along the chain.}

{
\begin{figure}[h!]
 \begin{center}
  \includegraphics[width=1\linewidth,clip,trim=0cm 0cm 0cm 0cm]{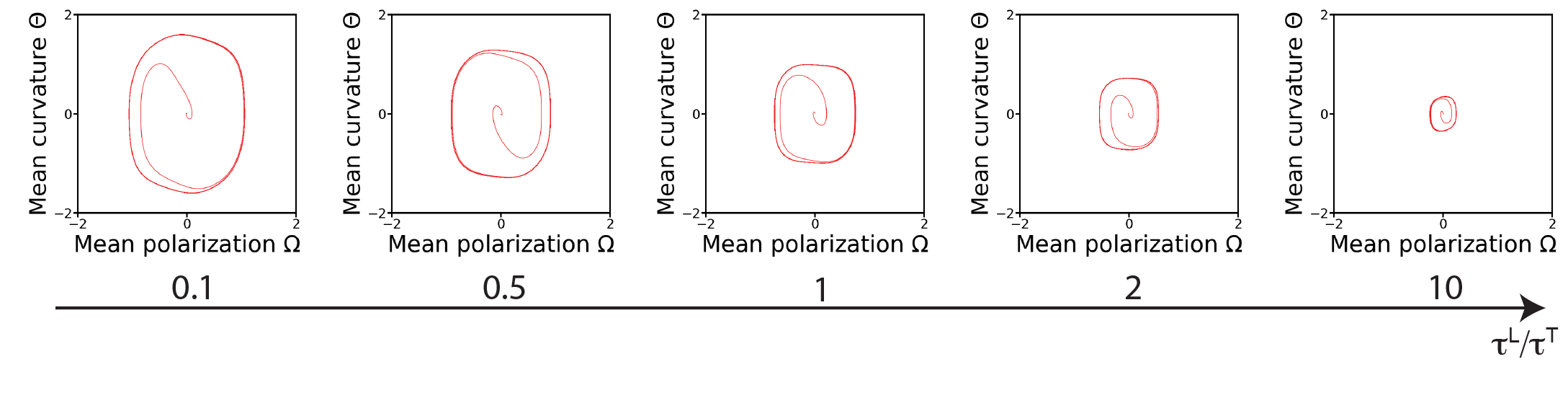}
 \end{center}
 \caption{{{\bf Role of anisotropy.} Curvature $\Theta=\theta_2-\theta_1$ vs. mean polarization $\Theta=(\theta_1+\theta_2)/2$ of two coupled pendulums with follower forces and an anisotropic drag for different values of the anisotropy ratio $\tau^L/\tau^T$. The other parameters were chosen as $\sigma=3.5$ and $\tau^T=1$.}}
\label{fig:anisotropy}
\end{figure}
}

\subsection{Derivation of the canonical form of the Hopf bifurcation}
In this section, we focus on the minimal case of an elasto-active chain with two active particles $N=2$, see Eqs.~(\ref{eq:sysN2a})-(\ref{eq:sysN2b}) of the Main Text and show that they can be mapped onto the Landau-Stuart equation, which is the normal form of a Hopf-bifurcation. First, we rewrite these equations as 
\begin{eqnarray}
\dot{\theta}_1&=&\frac{-\sigma  \sin (\theta_1-\theta_2)+(\theta_1-\theta_2) \cos (\theta_1-\theta_2)+2 \theta_1-\theta_2}{\tau  \left(\cos ^2(\theta_1-\theta_2)-2\right)}\\
\dot{\theta}_2&=&\frac{\sigma  \sin (2 \theta_1-2 \theta_2)+2 (\theta_2-2 \theta_1) \cos (\theta_1-\theta_2)-4 \theta_1+4 \theta_2}{\tau  (\cos (2 \theta_1-2 \theta_2)-3)}
\end{eqnarray}
We find that the Jacobian of this system of equation admits conjuguate eigenvalues, whose real part crosses zero at $\rho=3$. We therefore introduce the variable $\mu:=\rho-3$ and we will focus in the following on the transition point at $\mu=0$. In order to map onto the canonical form, we define the variable $z:=a\theta_1+i b\theta_2$, where $a$ and $b$ are arbitrary real numbers. To find which values of $a$ and $b$ allow up to find the canonical form, we first linearize the equation. We find
\begin{equation}
    \tau \dot{z}= \frac{(b+i a) (a (\mu +1)-i b (\mu -1))}{2 a b}z -\frac{i \left(a^2 (\mu +1)+b^2 (\mu -1)\right)}{2 a b}z^* +\mathcal{O}(z^2).
\end{equation}
The linear term of the canonical form of the Hopf-bifurcation has no $z^*$ term, therefore we choose $a$ and $b$ such that this term cancels out, that is $a=1$ and $b/a=\sqrt{(1+\mu)/(1-\mu)}$. We now use these values and rewrite $\dot{z}$ up to cubic order in $z$ and linearize the expression in $\mu$
\begin{equation}
\begin{split}
    \tau \dot{z}=&\quad 
     (i+\mu) z +\frac{(10+4 i) \mu +(-1+8 i)}{8}|z|^2 z
     \\
     &
     +\frac{-\left(1-2i \right) (16 \mu +8(2-3 i))}{8} |z|^2 z^*
     +\frac{(6-9 i)-(24+14 i) \mu}{24} z^3
     +\frac{(14+12 i) \mu +(6-9 i)}{24}  {z^*}^3
     \\
     &
    +\mathcal{O}(z^4).
    \end{split}
    \label{eq:Hopf_almost}
\end{equation}
Variable changes of the form $z=\tilde{z}+\alpha_{pq}\tilde{z}^p\tilde{z^*}^q$, with $p\in[0,3]$, $q\in[0,3]$ and $p+q=3$ can cancel out the cubic terms of Eq.~(\ref{eq:Hopf_almost}), except for $p=2$ and $q=1$, where the associated constraint diverges at the bifurcation $\mu=0$. Thereby, we obtain the final equation
\begin{equation}
    \tau \dot{z}=(i+\mu) z +\left(-1-\frac{\mu}{2} +i\left(-\frac{1}{8}+\frac{5}{4}\mu\right)\right) |z|^2 z+\mathcal{O}(z^4)
\label{eq:Hopf}.
\end{equation}
The prefactor of the cubic term is called the first Lyapunov coefficient and its real part is negative for small $\mu$, therefore the Hopf bifurcation is supercritical \cite{HolmesBook}, as discussed in the Main Text.

\subsection{Synchronization of two coupled elasto-active chains}
In this section, we investigate theoretically the nature of the synchronization transition observed in Fig. 4 of the Main Text. To this end, we focus on with the minimal case of two elastically coupled elasto-active chains. Following Eqs.~(\ref{eq:motion_d1}-\ref{eq:motion_d1N}), such elasto-active chains are described by the following set of ODEs:
\begin{eqnarray}
\dot{\theta}_1&=&-\frac{-\theta _1 \left(\cos \left(\theta _1-\theta _2\right) +2\right)+\sigma  \sin \left(\theta _1-\theta _2\right)+\theta _2 \left(\cos \left(\theta _1-\theta _2\right)+1\right)-\kappa (\theta_1 - \phi _1)}{\tau  \left(\cos ^2\left(\theta _1-\theta _2\right)-2\right)}\\
\dot{\theta}_2&=&\frac{2 \cos \left(\theta _1-\theta _2\right) \left(\theta _2+\kappa  \phi _1\right)-2 \theta _1 \left((\kappa +2) \cos \left(\theta _1-\theta _2\right)+2\right)+\sigma  \sin \left(2 \left(\theta _1-\theta _2\right)\right)+4 \theta _2}{\tau  \left(\cos \left(2 \left(\theta _1-\theta _2\right)\right)-3\right)},\\
\dot{\phi}_1&=&
-\frac{\theta _1 \kappa -\phi _1 \left(\kappa +\rho  \cos \left(\phi _1-\phi _2\right)+2 \rho \right)+\rho  \phi _2 \left(\cos \left(\phi _1-\phi _2\right)+1\right)+\sigma  \sin \left(\phi _1-\phi _2\right)}{\tau  \left(\cos ^2\left(\phi _1-\phi _2\right)-2\right)}
\\
\dot{\phi}_2&=&\frac{-2 \cos \left(\phi _1-\phi _2\right) \left(\phi _1 (\kappa +2 \rho )-\theta _1 \kappa \right)-4 \rho  \phi _1+2 \rho  \phi _2 \left(\cos \left(\phi _1-\phi _2\right)+2\right)+\sigma  \sin \left(2 \left(\phi _1-\phi _2\right)\right)}{\tau  \left(\cos \left(2 \left(\phi _1-\phi _2\right)\right)-3\right)}.
\end{eqnarray}
Based on the analysis carried out above on a single chain, we change variables $z_1:=\theta_1+a_1 i \theta_2$, where $a_1=\sqrt{(1+\mu)/(1-\mu)}$ and $z_2:=\phi_1+a_2 i \phi_2$, where $a_2=\sqrt{(3+\mu-2\rho)/(-3-\mu+4\rho)}$ and then to linear order in $\mu$ and $\rho-1$. Such limits correspond to the vicinity of the bifurcation $\mu\ll 1$ and that the case where the two chains have almost the same elasticity $\rho-1\ll 1$. Such hypothesis is not strictly necessary to proceed and does not affect the spirit of the following derivation, yet it drastically simplifies the algebraic manipulations.
As a result, we obtain the two coupled ODEs:
\begin{eqnarray}
\dot{z}_1&=& (i+\mu) z_1 
+\left(-1-\frac{\mu}{2} +i\left(-\frac{1}{8}+\frac{5}{4}\mu\right)\right) |z_1|^2 z_1 
+\frac{\kappa}{2}(i(\mu+1)-1)(z_1+z_1^\dagger-z_2+z_2^\dagger)\\
\dot{z}_2&=&
\left(i (3 \mu  \rho -3 \mu +\rho )+\mu -3 \rho +3\right)z_2 + \left(\frac{i}{8}\left(- 9 \mu  \rho +19 \mu-31 \rho +30\right)-3 \mu  \rho +\frac{5 \mu }{2}+\frac{\rho }{2}-\frac{3}{2}\right) |z_2|^2 z_2\nonumber\\ &&+\frac{\kappa}{2}(i (-4 \mu  \rho +5 \mu -3 \rho +4)-1)(z_2+z_2^\dagger-z_1+z_1^\dagger)
\end{eqnarray}
These two coupled ODEs represent two coupled oscillators close to a Hopf bifurcation. To proceed further, we assume that the coupling between the chains $\kappa\ll 1$. We can then study the coupling as a perturbation about the two limit cycles of the two elasto-active chains~\cite{niedermayer2008synchronization}. This hypothesis relies on the assumption that the coupling does not affect the magnitude the self-oscillations, but that it affect their phase. In other words, if we introduce the amplitude and phase of the complex variables $z_1$ and $z_2$ by introducing the following variable changes
$z_1=R_1 e^{i\Phi_1}$ and $z_2=R_2 e^{i\Phi_2}$, the amplitudes will be given by the uncoupled chains and with remain constant 
$R_1=\sqrt{\mu}$ and $R_2=\sqrt{\mu -3 (\rho -1)}$, 
and the time-evolution of the phases will be given by the following equations
\begin{eqnarray}
\frac{d \Phi_1}{dt}&=& 1-\frac{\mu}{8}
+\frac{\kappa}{2}\left[1+\frac{\sqrt{\mu -3 \rho +3}}{\sqrt{\mu }}(\sin(\Phi_2-\Phi_1)-\cos(\Phi_2-\Phi_1))\right]\\
\frac{d \Phi_2}{dt}&=&1-\frac{1}{8}(\mu+11(1-\rho))
+\frac{\kappa}{2}\left[1-\frac{\sqrt{\mu }}{\sqrt{\mu -3 \rho +3}}(\sin(\Phi_1-\Phi_2)-\cos(\Phi_1-\Phi_2))
\right],
\end{eqnarray}
which is linearized with respect to $\mu$, $\rho-1$ and $\kappa$ and where the non-resonant terms proportional to $\sin(2\Phi_1), \sin(2\Phi_2)$ and $\sin(\Phi_1+\Phi_2)$, on the right-hand side have been neglected. Such terms typically average out and do not contribute to changing the relative phase between the two oscillators $\Phi_1+\Phi_2$~\cite{niedermayer2008synchronization}. Finally, we can subtract the last two equations to express the equation governing the time-evolution of the instantaneous phase difference
between the two chains $\Psi:=\Phi_2-\Phi_1$ to obtain
\begin{equation}
    \frac{d\Psi}{dt}=d\nu+\varepsilon(\sin\Psi + \tan\Psi_0\cos\Psi),
\end{equation}
where $d\nu:=\frac{11}{8}(\rho-1)$, $\varepsilon:= \frac{\kappa  (2 \mu -3 \rho +3)}{2 \sqrt{\mu } \sqrt{\mu -3 \rho +3}}$ and $\tan\Psi_0:=\frac{3 (\rho -1)}{2 \mu -3 \rho +3}$. This equation can be recast as 
\begin{equation}
    \frac{d\Psi}{dt}=d\nu+\frac{\varepsilon}{\cos\Psi_0}\sin\Psi -\Psi_0,
\end{equation}
which is a standard equation for non-isosynchronous synchronization~\cite{niedermayer2008synchronization}. The synchronization occurs when $|d\nu|<|\varepsilon/\cos\Psi_0|$, which, translated in the parameters of the problem, corresponds to coupling constants $\kappa$ larger than $\kappa_c=11/(4\sqrt{6}) \sqrt{\mu |1-\rho |}$. Synchronization occurs for vanishingly small coupling close to the bifurcation of the first chain $\mu=0$ and in the limit where the two chains are identical, that is $\rho=1$, as can be seen in the inset of Fig.~4f of the Main Text.

\rev{
\subsection{Chain pinned at both ends}
\subsubsection{Derivation of the equations of motion}
Repeating the derivation above with the constraints 
\begin{equation}
    \sum_{i=1}^N\cos\theta_i=N(1-\delta),
    \label{eq:constraint1}
\end{equation}
where $\delta$ is the amount by which the chain is compressed along its axis and
\begin{equation}
   \sum_{i=1}^N\sin\theta_i=0,
   \label{eq:constraint2}
\end{equation}
we find
\begin{eqnarray}
0&=&2\theta_1\!-\!\theta_2 \!-\! \sigma\sum_{j=1}^N\!\!\sin(\theta_1\!-\!\theta_j) \!+\! \tau  \!\left(\!N\dot{\theta}_1+\sum_{j=2}^N \dot{\theta}_j\!\cos(\theta_1\!-\!\theta_j)\!\right)-\mu\sin\theta_1 + \eta \cos\theta_1 \textrm{  for }i\!=1\label{eq:motion_1}\\
\nonumber 0&=& 2\theta_i\!-\!\theta_{i-1}\!-\!\theta_{i+1} \!-\! \sigma\sum_{j=i}^N\sin(\theta_i\!-\!\theta_j)\\
&&+\tau \!\left(\!\!(\!N\!-\!i\!+\!1\!)\!\sum_{j=1}^i\!\dot{\theta}_j\!\cos(\theta_i\!-\!\theta_j)\!+\!\sum_{j=i+1}^N \!(\!N\!-\!j\!+\!1\!)\!\dot{\theta}_j\!\cos(\theta_i\!-\!\theta_j)\!\right) \\
&&-\mu\sin\theta_i + \eta \cos\theta_i \textrm{  for }i\!\in\![2,N-1]\label{eq:motion_i}\\
0&=& \theta_N\!-\!\theta_{N-1} \!+\! \tau\sum_{j=i}^N \!\dot{\theta}_j\!\cos\left(\theta_N\!-\!\theta_j\right) -\mu\sin\theta_N + \eta \cos\theta_N\textrm{  for }i\!=N,\label{eq:motion_N}
\end{eqnarray}
where $\mu$ and $\eta$ are Lagrange multipliers enforcing the constraints Eq.~\eqref{eq:constraint1} and Eq.~\eqref{eq:constraint2}, respectively.
We solve this system of differential algebraic equation (DAE) again using Mathematica, by reducing the step size and enforcing a time constraint on the equation simplification, since the additional constraints render the DAE stiff.
In the Main Text, we restrict our attention to the case $\sigma=0.6$ that corresponds to the experimental case. Below in Fig.~\ref{fig:phase_doublepinned} for a chain with $N=9$, we show a plot the magnitude of oscillation as a function of both the compression $\delta$ and of the elasto-active number $\sigma$. We see as in the Main Text that that the larger $\delta$, the larger the oscillations. We also see that only for large enough elasto-active number does the self-snapping occurs.}

\begin{figure}[h!]
 \begin{center}
  \includegraphics[width=0.4\linewidth,clip,trim=0cm 0cm 0cm 0cm]{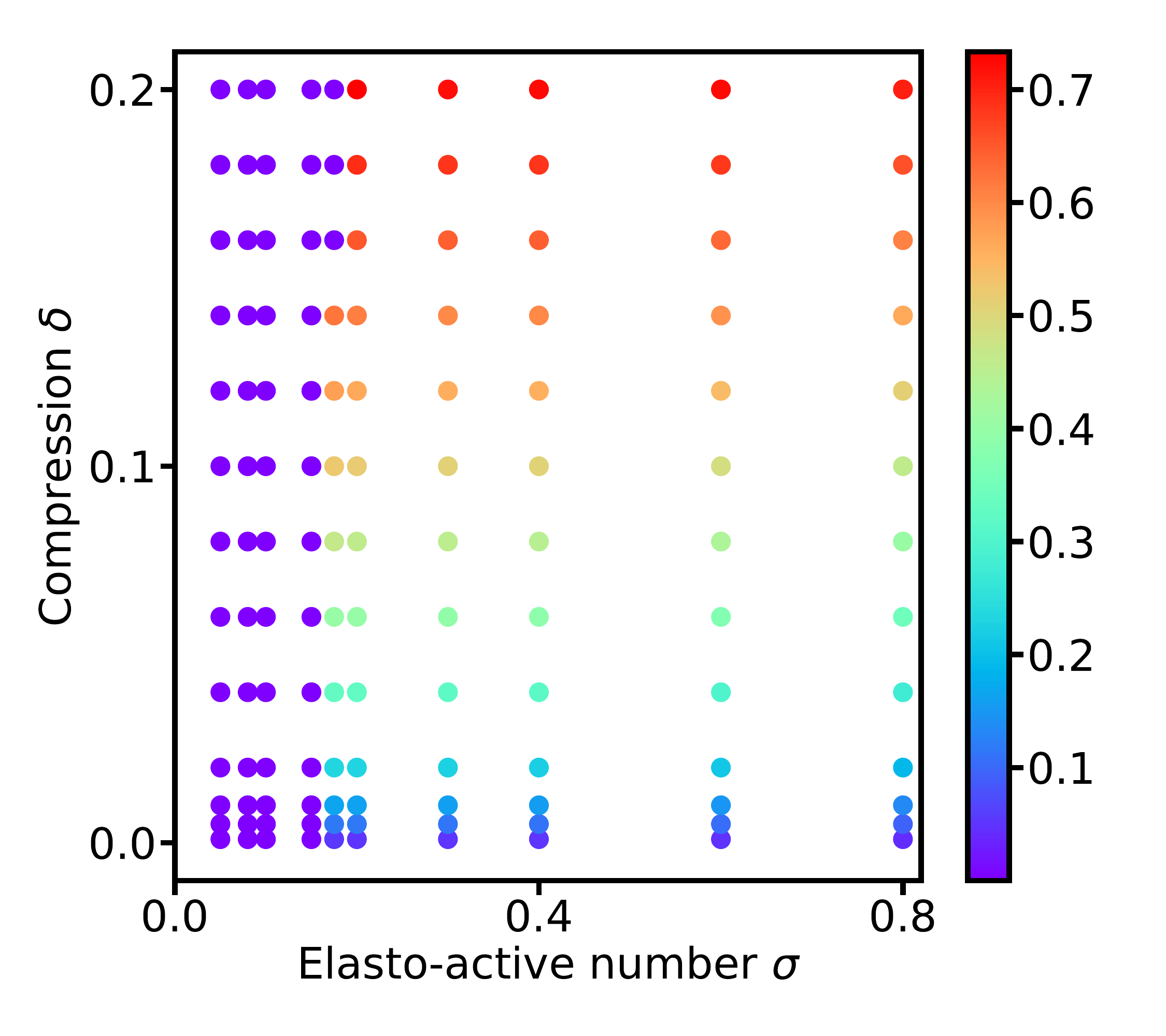}
 \end{center}
 \caption{\rev{{{\bf Self-snapping.} The color-map denotes the amplitude (in radians) of self-snapping measured by the standard deviation of the third particle from the pinning point as a function of the the compression $\delta$ and of the elasto-active number $\sigma$ of the elasto-active chain.}}}
\label{fig:phase_doublepinned}
\end{figure}

\rev{
\subsubsection{Linear limit}
Eqs.~(\ref{eq:motion_1})-(\ref{eq:motion_N}) are impossible to solve explicitly in their full generality. We will see below that unlike the case considered above of a single pinning point, even linear stability is difficult in this case, because one does not know around which state to linearize. Imagine we would linear close to the limit where all the angle are zero. 
\begin{eqnarray}
0&=& 2\theta_1\!-\!\theta_2 \!-\! \frac{Fa^2}{C}\sum_{j=1}^N\!\!(\theta_1\!-\!\theta_j) \!+\! \frac{\gamma a^2}{C} \!\left(\!N\dot{\theta}_1+\sum_{j=2}^N \dot{\theta}_j\!\right)-\mu\theta_i+\eta \label{eq:motion_1_lin}\\
0&=& 2\theta_i\!-\!\theta_{i-1}\!-\!\theta_{i+1} \!-\! \frac{Fa^2}{C}\sum_{j=i}^N(\theta_i\!-\!\theta_j)\!+\! \frac{\gamma a^2}{C} \!\left(\!\!(\!N\!-\!i\!+\!1\!)\!\sum_{j=1}^i\!\dot{\theta}_j\!\!+\!\sum_{j=i+1}^N \!(\!N\!-\!j\!+\!1\!)\!\dot{\theta}_j\!\right)-\mu\theta_i+\eta\textrm{ for }i\!\in\![2,N-1] \label{eq:motion_i_lin}\\
0&=& \theta_N\!-\!\theta_{N-1} \!+\! \frac{\gamma a^2}{C} \sum_{j=i}^N \!\dot{\theta}_j-\mu\theta_N+\eta.\label{eq:motion_N_lin}\\
N(1-\delta)&=& \sum_{j=1}^N 1  \\
0&=& \sum_{j=1}^N\theta_i \\
\end{eqnarray}
The stability of the system can in principle be investigated by injecting the following ansatz $(\theta_1,\theta_2,\dots,\theta_N)=(\theta^0_1,\theta^0_2,\dots,\theta^0_N,\mu,\eta)\exp\lambda t$
in Eqs.~(\ref{eq:motion_1_lin})-(\ref{eq:motion_N_lin}) and requiring that the determinant of such system of equation to be zero, which which can be rewritten in the following matrix form:
$A_c.(\theta^0_1,\theta^0_2,\dots,\theta^0_N,\mu,\eta)^T=(0,0,\dots,0,-N\delta,0)^T$, with 
\begin{equation}
    A_c=\left(\begin{array}{cc}
        A & M^T \\
        M & 0
    \end{array}\right),
\end{equation}
where $M = \left(\begin{array}{cccccc}
        0&0&0&0&\cdots&0\\
        1&1&1&1&\cdots&1
    \end{array}\right)$ and 
\begin{equation}
A_{ij}=(2\delta_{ij}-\delta_{i,j-1}-\delta_{i-1,j})+ \left((N-j+1) u_{ij} + (N-i+1) d_{ij}\right)\tau\lambda +\left(u_{ij}-(N-i) \delta_{ij}\right)\sigma,
\end{equation}
where $\delta_{ij}$ is the Kronecker delta, $u_{ij}$ the strictly upper triangular matrix and $d_{ij}$ the lower triangular matrix. By solving det$A=0$, we should be able to find the eigenvalues $\lambda$ at a function of the elasto-active number $\sigma$ and of the compression $\delta$. 
However, the only situation that is kinematically admissible for $\theta=0$ correspond to the case where the chain is straight, i.e. when $\delta=0$. This provide limited information about the case where the chain self-snaps unfortunately, so we don't use it in the Main Text.}

\bibliography{biblio.bib}